\documentclass{JHEP3}
\usepackage{amsmath,amssymb}
\usepackage{graphicx}

\newcommand{\beq}{\begin{eqnarray}}
\newcommand{\eeq}{\end{eqnarray}}
\newcommand{\half}{\frac12}

\newcommand{\wg}{\wedge}

\def\O{{\mathcal{O}}}

\def\cF{{\mathcal{F}}}

\def\cK{{\mathcal{K}}}

\def\cM{{\mathcal{M}}}

\def\bz{\bar{z}}

\def\be{\beta}

\def\nn{\nonumber}

\def\r2{{\sqrt{2}}}
\def\h{{\eta}}

\def\h0{\hat{h}}
\def\Vr0{\hat{V}_{r}}
\def\Vp0{\hat{V}_{\phi}}

\def\r2{\sqrt{2}}
\def\be{\begin{equation}}
\def\ee{\end{equation}}
\def\nn{\nonumber}

\def\tq{\tilde{q}}
\def\hJ{\hat{J}}
\def\hcF{\hat{\mathcal{F}}}

\def\hB{\hat{B}}
\def\heta{\hat{\eta}}

\def\si1{\sin\theta_1}
\def\si2{\sin\theta_2}
\def\cs1{\cos\theta_1}
\def\cs2{\cos\theta_2}
\def\T21{\tan^2\theta_1/2}
\def\T22{\tan^2\theta_2/2}
\def\CT21{\cot^2\theta_1/2}
\def\CT21{\cot^2\theta_1/2}

\title{On Supersymmetric D7-branes in the Warped Deformed Conifold}

\author{Heng-Yu Chen$^1$, Peter Ouyang$^1$, and Gary Shiu$^{1,2}$ \\
$^1$ Department of Physics, University of Wisconsin, Madison,
WI 53706, USA
\vskip 3pt
$^2$ PH-TH Division, CERN, CH-1211 Geneva 23, Switzerland}

\date{\today}

\abstract{We study the supersymmetric properties of D7-branes in the warped deformed conifold.
We consider the $\kappa$-symmetry conditions on D7-branes in this
specific warped background, taking into account the background NS-NS
2-form flux. While any holomorphic embedding defines a
supersymmetric D7-brane in the absence of background H-flux, most of the
D7-brane embeddings considered in the literature do not preserve
supersymmetry for the warped deformed conifold without also including brane
worldvolume flux.  For the simplest such embedding, we construct numerically the worldvolume flux
necessary to restore supersymmetry. We also
comment on the dual field theory descriptions in terms of  cascading
${\cal N}=1$ supersymmetric gauge theories with flavors. Finally, we
discuss some possible applications of our results to moduli stabilization
and vacuum energy uplifting, gauge/gravity duality, and string
inflationary model building.}


\preprint{MAD-TH-08-11\\
CERN-PH-TH/2008-150}

\begin{document}

\section{Introduction}
\paragraph{}

One of the major developments in string theory of the last decade
has been the emergence of warped spacetimes as a setting for a rich
variety of physics.  To date, perhaps the most important application
of such warped geometries has been in the gauge theory/string theory
correspondence
\cite{jthroat,US,EW}. In this setting, noncompact spacetimes which
are warped by the presence of background flux have been conjectured
to give a dual description of gauge theories.  In many cases, this
duality can relate strongly coupled field theories to weakly coupled
gravity theories, and as such offers new tools for studying field
theories when traditional perturbative methods fail.

Another application of warped geometry has been in the study of
string theory as a theory of particle physics.  When one
compactifies string theory from ten dimensions to four dimensions,
the compactification manifold can support nontrivial background
fluxes, and these fluxes will generically have a backreaction on the
geometry, resulting in warping.
The gravitational redshift generated by a strong warp factor can
lead to a hierarchy of scales, thus offers a geometrical explanation
for the huge disparity between the electroweak and the Planck scale
\cite{RS} (see also \cite{Verlinde,Sethi,Shiu,GKP} for string theory realizations).
For the same reason, warped geometries have also been invoked in
supersymmetry breaking scenarios in string theory, both to lower the
scale of supersymmetry breaking \cite{Burgess:2006mn,Douglas:2007tu}
and to control the amount of sequestering \cite{warpedsequestering}.
The low energy effective supergravity theory describing strongly warped backgrounds is challenging to derive, though continued progress has been made \cite{Burgess:2006mn,Douglas:2007tu,DeWolfe:2002nn,Giddings:2005ff,Frey:2006wv,Shiu:2008ry,Douglas:2008jx}.

Warping has also found application in inflationary cosmology.
A key issue in inflationary cosmology is to understand the
underlying dynamics which allows for a sufficient number e-folds of
expansion.  It turns out that warping can be fruitfully used in
stringy cosmological models to construct suitable potentials for
slow roll inflation
as well as for motivating new inflationary mechanisms
(see
\cite{stringcosmologyreviews} and the references therein).

Strikingly, there is a particular warped geometry, the warped
deformed conifold in type IIB supergravity \cite{Conifold,KS}, which
is simple enough to allow detailed study, but is also rich enough to
have been used in all three of these areas.  It has almost anti-de
Sitter asymptotics, and in gauge/gravity duality it is dual to an
interesting confining gauge theory.  If one embeds the deformed
conifold in a compact Calabi-Yau, and turns on appropriate
three-form fluxes, one finds that the resulting background has a
natural and stable hierarchy of scales \cite{GKP}.  And in
cosmology, interesting inflationary models have been also
constructed in the conifold
\cite{KKLMMT,Baumann:2006th,DelicateU,Krause:2007jk}.

In this paper, we study a particularly interesting variation of the
warped deformed conifold, in which we add a certain number of probe
D7-branes to the background.  These branes fill four noncompact
directions and wrap a four-cycle $\Sigma_4$ in the transverse space.
In the gauge theory dual, these branes add fundamental matter to the
theory \cite{Aharony:1998xz,Karch}, offering the prospect of field
theory duals which resemble QCD.  Moreover, in more phenomenological
applications, the D7-branes allow the introduction of gaugino
condensates, which can stabilize K\"{a}hler moduli \cite{KKLT} and
which can introduce potentials for mobile D3-branes (candidate inflatons in
the scenarios considered in \cite{KKLMMT,Baumann:2006th,DelicateU}).
Given
this range of applications, it is of clear interest to find as many
supersymmetric D7-brane embeddings in the deformed conifold as
possible, allowing for greater freedom in model-building. In this
paper we study the supersymmetry conditions for several different
classes of branes, focusing in particular on the cases where
worldvolume flux is necessary for the brane to preserve
supersymmetry.

This paper is organized as follows.
We begin by reviewing the deformed conifold in Section
\ref{ReviewDefcon}, along the way collecting some results which will
be useful in
the rest of the paper.  In Section
\ref{SUSYembed}, we
review the criteria for the D7-branes to preserve supersymmetry (the
$\kappa$-symmetry conditions) in
a general Calabi-Yau flux compactification. We make use of these
results in Section \ref{SysSUSYD7}, where we study the case of
D7-branes in the conifold with background fluxes. There is one such
D7-brane embedding (sometimes known as the Kuperstein embedding)
already known to be supersymmetric with no worldvolume flux turned
on \cite{Kuper}. On the other hand, the embedding first studied in
\cite{Ouyang} is not supersymmetric when the worldvolume flux vanishes.
Our main result is in Section 4.3, where we construct the
worldvolume flux necessary to restore supersymmetry, by a
combination of analytic and numerical methods.  We also offer a
criterion that in many cases can show whether a worldvolume flux is
needed to restore supersymmetry, and which is far simpler to verify
than explicitly checking $\kappa$-symmetry. In Section
\ref{FieldTheory},
we make some observations about the field theories dual to the
conifold with our D7-branes included.  Finally, we end with a
discussion of prospects for the future in Section
\ref{Conclusion}.

\section{Review of the Warped Deformed Conifold}\label{ReviewDefcon}
\paragraph{}

In this paper we focus on a particular concrete example of a
warped Calabi-Yau space,
namely the warped deformed conifold.
This geometry is noncompact, but in principle it can be embedded in
a compact Calabi-Yau with background bulk fluxes.  We now proceed to
review some facts about the conifold and warped compactifications in
general which will be useful in the rest of the paper.

The conifold is most simply defined as a submanifold of flat ${\mathbb{C}}^4$
given by the equation \cite{Conifold}
\beq
z_1^2+z_2^2+z_3^2+z_4^2 = 0
\label{coneq}
\eeq
where the $z_i$ are coordinates of the ambient ${\mathbb{C}}^4$.  This space
has a manifest $SU(2)\times SU(2) \times U(1)$ global symmetry,
where the $SU(2)\times SU(2) \simeq SO(4)$ acts by rotating the
$z_i$ and the $U(1)_R$ acts as multiplication of each of the $z_i$
by a phase, $z_i \rightarrow e^{i\varphi} z_i$. The defining
equation of the conifold is also invariant under an overall
rescaling of the coordinates, which implies that it admits a conical
metric. The corresponding geometry is a cone over a five-dimensional
Einstein manifold called $T^{1,1}$ and the metric takes the form
\beq
ds_6^2 = dr^2 +r^2 ds_{T^{1,1}}^2.
\eeq
where the metric of the $T^{1,1}$ base is
\beq \label{co}
ds_{T^{1,1}}^2= {1\over 9} \bigg(d\psi + \sum_{i=1}^2 \cos \theta_i
d\phi_i\bigg)^2+ {1\over 6} \sum_{i=1}^2 \left( d\theta_i^2 + {\rm
sin}^2\theta_i d\phi_i^2
 \right)
\eeq
and the angular coordinates range as $0\le \psi \le 4\pi$, $0\le
\theta_i \le \pi$, and $0 \le \phi_i \le 2\pi$.
From the form of the metric it is clear that base is a sphere
fibration $S^1\times S^2\times S^2$ and it turns out that its
topology is actually $S^3
\times S^2$. The base space has Betti numbers $b_2=b_3=1$, and the
associated harmonic forms are
\beq
\omega_2 &=& \half \left( \Omega_{11} -\Omega_{22}\right) \\
\omega_3 &=& \zeta \wg \omega_2
\eeq
where $\zeta=d\psi+\cos\theta_1 d\phi_1 +\cos\theta_2 d\phi_2$ and
\beq
\Omega_{ij} = d\theta_i \wg \sin\theta_j d\phi_j.
\label{Omega}
\eeq
This space is singular at the tip of the cone where all the $z_i$
are zero (and so we will sometimes refer to this space as the
``singular conifold.")

It is also possible to define the conifold in terms of another set
of complex coordinates, by the equation
\beq
w_1 w_2 - w_3 w_4 = 0
\label{conifw}
\eeq
which is related to the original defining equation by an obvious
change of variables.  It turns out that this alternative description
will also be useful to us.  One nice property is that the $w_i$
coordinates admit a relatively simple description in terms of the
angles on $T^{1,1}$:
\beq
w_1 &=& r^{3/2} e^{i/2(\psi-\phi_1-\phi_2)}\sin\frac{\theta_1}{2}\sin\frac{\theta_2}{2}, \\
w_2 &=& r^{3/2} e^{i/2(\psi+\phi_1+\phi_2)}\cos\frac{\theta_1}{2}\cos\frac{\theta_2}{2}, \\
w_3 &=& r^{3/2}e^{i/2(\psi-\phi_1+\phi_2)}\sin\frac{\theta_1}{2}\cos\frac{\theta_2}{2}, \\
w_4 &=& r^{3/2}
e^{i/2(\psi+\phi_1-\phi_2)}\cos\frac{\theta_1}{2}\sin\frac{\theta_2}{2}.
\label{wtoangles} \eeq

One of the important properties of the conifold is that its
singularity can be smoothed by adjusting moduli.  At the tip of the
cone, one may perform either a small resolution, blowing up an
$S^2$, or a deformation, by changing the defining equation of the
conifold to
\beq
z_1^2+z_2^2+z_3^2+z_4^2 = \epsilon^2.
\label{DefCon}
\eeq
We can make a phase rotation of the $z_i$ so that $\epsilon$ is
real, and we will always assume this in the rest of the paper.  It
is evident from taking a real slice of this defining equation that
deformation results in the appearance of a finite-sized $S^3$ at the
tip of the cone.  The deformed conifold preserves the $SO(4)$
isometry of the singular conifold, but the scale invariance is
broken by $\epsilon$ and the $U(1)_R$ symmetry is broken to a ${\mathbb{Z}}_2$
subgroup.

This deformed geometry has special relevance in flux
compactifications because of its topology.  The existence of a
three-cycle allows us to turn on RR 3-form flux on this cycle, and
there are supersymmetric supergravity solutions if we also include
NS-NS flux on the (noncompact) dual cycle.

We now proceed to review flux compactification of the type
considered in
\cite{GKP}, where a warped deformed conifold throat naturally
develops near the local conifold singularities. Let us begin with
the ten dimensional warped metric preserving four-dimensional
Poincare symmetry which takes the following form
\begin{equation}
ds^2_{10}=e^{2A(y)}\tilde{g}_{\mu\nu}dx^{\mu}dx^{\nu}+e^{-2A(y)}\tilde{g}_{mn}dy^m dy^n\label{10warpedmetric}\,.
\end{equation}
Here $\tilde{g}_{\mu\nu}\,,\mu,\nu=0,1,2,3$ and $\tilde{g}_{mn}\,,m,n=4\,,\dots\,, 9$ are the unwarped metrics for the non-compact four dimensional space time and the compact six dimensional manifold ${\mathcal{M}}_6$ respectively, whereas $e^{A(y)}$ is the warp factor which only varies over ${\mathcal{M}}_6$.

Following \cite{GKP}, we shall allow for both RR and NS-NS fluxes of
IIB supergravity to be turned on, and they can be written succinctly
as:
\begin{eqnarray}
G_3&=&F_3-\tau H_3=\frac{1}{6}G_{mnp}[dy^m\wedge dy^n\wedge dy^p]\,,\label{DefG3}\\
\tilde{F}_5&=&F_5-\frac{1}{2}C_2\wedge H_3+\frac{1}{2}B_2\wedge F_3=(1+\ast_{10})[d\alpha(y)\wedge dx^0\wedge dx^1\wedge dx^2 \wedge dx^3]\label{DefTF5}\,.
\end{eqnarray}
Here we have allowed for IIB axio-dilaton $\tau\equiv
\tau(y)=C_0-ie^{-\phi}$ to vary over $\cM_6$ and in (\ref{DefTF5})
we have used self-duality $\tilde{F}_5=\ast_{10}\tilde{F}_5$, four
dimensional Poincare invariance and Bianchi identity to constrain
the expression for $\tilde{F}_5$. In addition, local objects
extending into non-compact four dimensions can also be included
(which may wrap cycles in $\cM_6$), and they need to satisfy the
tadpole cancellation condition:
\begin{equation}
\frac{1}{2\kappa_{10}^2T_3}\int_{\cM_6} H_3\wedge F_3+Q_3^{\rm local}=0\,,\label{Tadpolecond}
\end{equation}
where $Q_3^{local}$ is the D3-brane charge on the local objects.
The supergravity equation of motion for such configuration of fluxes and local sources then yields:
\begin{equation}
\tilde{\nabla}^2\Phi_{\pm}=\frac{e^A}{6 {\rm Im}(\tau)}|G_{\pm}|^2+|\nabla \Phi_{\pm}|^2+2\kappa_{10}^2 e^{2A}\left[\frac{(T^{m}_{m}-T^{\mu}_{\mu})^{\rm local}}{4}-T_3\rho_3^{\rm local}\right]\label{SUGRAeom}\,,
\end{equation}
where we have defined the following linear combination:
\begin{equation}
\Phi_\pm= e^{4A(y)}\pm \alpha(y)\,,~~~~G_\pm=iG_3\pm \ast_6 G_3\,,\label{DefPhiGpm}
\end{equation}
and $T_{\mu\nu}$ and $T_{mn}$ are four and six dimensional energy-momentum tensors. A special solution to (\ref{SUGRAeom}) is given by the so-called imaginary self-dual (ISD) configuration:
\begin{equation}
\Phi_{-}=0\,,~~G_{-}=0\,,~~\leftrightarrow~~ e^{4A(y)}=\alpha(y)\,,~~\ast_6 G_3=iG_3\,,\label{ISDconfig}
\end{equation}
\begin{equation}
\frac{(T^{m}_{m}-T^{\mu}_{\mu})^{\rm local}}{4}=T_3\rho_3^{\rm local}\,.\label{ISDconfig2}
\end{equation}
In terms of the local sources in IIB, the equality (\ref{ISDconfig2}) can be satisfied by D3 branes or O3 planes, D5 brane wraps on vanishing two cycle, also known as ``fractional D3''; or D7 wraps on non-trivial four cycles, e.g. K3.
Turning now from these generalities back to the conifold, let us
turn on $M$ units of $F_3$ along the so-called A-cycle which
corresponds to the minimal $S^3$ in the deformed conifold, and also
$(-K)$ units of $H_3$ in the dual B-cycle.
The quantization condition for the fluxes then yields:
\begin{equation}
\frac{1}{2\pi\alpha '}\int_A F_3=M\,,~~~ \frac{1}{2\pi\alpha '}\int_B H_3=-K\,.\label{3formquantizations}
\end{equation}
These three form fluxes stabilize the value of $\epsilon^2$ at
\begin{equation}
\epsilon^{2}\sim \exp(-2\pi K/g_s M)\,,\label{Stabilizedepsilon}
\end{equation}
and the warp factor at the tip is given by $e^{A_0}=e^{-2\pi
K/3Mg_s}$. The total D3 charge is given by the intersection form:
\begin{equation}
\frac{1}{2\kappa_{10}^2 T_3}\int_{\cM_6}H_3\wedge F_3=MK=N\,.\label{D3charges}
\end{equation}
The explicit IIB supergravity solution corresponds to the flux configuration described here was given in \cite{KS}, where the authors started with $N$ unit of D3 branes and $M$ unit of fractional D3 branes at the tip of singular conifold, with $g_s N, g_s M\gg 1$. In the near horizon limit, these branes are replaced by their fluxes, one can obtain explicit expressions for the corresponding warp factor and fluxes (for reviews, see e.g. \cite{ConNotes}, \cite{ConNotes2}).

In the large radius limit, the supergravity solution including these
3-form fluxes (but with the axion and dilaton constant) has the
useful description (due to Klebanov and Tseytlin, or KT) \cite{KT}:
\beq
ds_{10}^2 &=&  e^{2A(r)}(dx_{\mu} dx^{\mu}) + e^{-2A(r)}\left(dr^2+ r^2 ds_{T^{1,1}}^2\right) \label{ktmetric}\\
e^{-4A(y)} &=& \frac{27}{4r^4} \pi g_s\alpha'^2 \left( N + \frac{3}{2\pi} g_s M^2 \log(r/r_0)\right)\label{ktwarp}\\
g_s \tilde{F}_5 &=& d^4x \wg de^{4A(r)} + \ast_{10} ( d^4x \wg de^{4A(r)})\label{ktf5}\\
F_3 &=& \frac{M\alpha'}{2}\omega_3 \label{ktf3}\\
B_2 &=& \frac{3g_s M\alpha'}{2} \log(r/r_0) \wg \omega_2\\
H_3 &=& dB_2= \frac{3g_s M\alpha'}{2} \frac{dr}{r} \wg \omega_2.
\label{kth3}
\eeq
This solution possesses a naked singularity at small $r$; however,
in the full KS solution the deformation of the conifold resolves the
singularity.

For later use, we will record here the explicit K\"ahler form and
the NS-NS B-field on the warped deformed conifold.
$SO(4)$ invariant expressions can be found in
\cite{ConNotes}:
\begin{eqnarray}
J&=&\cK'(\rho)\left(\sum_{i=1}^4 dz_{i}\wedge d\bz_{i}\right)+\cK''(\rho)\left(\sum_{i=1}^{4}\bz_i dz_i\right)\wedge \left(\sum_{j=1}^{4} z_j d\bz_j\right)\label{DefJ}\,,\\
B_2&=&b(\rho)\epsilon_{ijkl} z_i\bz_j d z_k\wedge d \bz_l\,,~~~~
b(\rho)=\frac{ig_s M\alpha '}{2|\epsilon|^{4}}\frac{\rho\coth\rho-1}{\sinh^2\rho}\label{DefB}\,,
\end{eqnarray}
where $\cK(\rho)$ the K\"ahler potential for the deformed conifold
is defined implicitly as:
\begin{eqnarray}
\cK'(\rho)&=&\left(\frac{\sinh( 2\rho)-2\rho}{2|\epsilon|^2\sinh^3\rho}\right)^{\frac{1}{3}}\label{DefKahPot1}\,,\\
\cK''(\rho)&=&\frac{(5+\cosh(2\rho)-6\rho\coth\rho)}{6|\epsilon|^2\sinh\rho(\rho-\cosh\rho\sinh\rho)}\cK'(\rho)\,.\label{DefKahPot2}
\end{eqnarray}
Here ${()}^{\prime}$ denotes the differentiation with respect to
$r^3$, and the usual radial coordinate $r$
is related to the coordinate $\rho$ by
\begin{equation}
r^3=\sum^{4}_{i=1}|z_{i}|^2=|\epsilon|^2\cosh\rho\label{radialcoord}\,.
\end{equation}
The following definitions, taken from \cite{ConNotes2}, prove to be
notationally convenient:
\begin{eqnarray}
\eta_1=\epsilon_{ijkl} z_i\bz_j d z_k\wedge d \bz_l\,,~~~\eta_4=\sum_{i=1}^{4}\bz_i dz_i\wedge \sum_{j=1}^{4} z_j d\bz_j\,,~~~\eta_5=\sum_{i=1}^4 dz_{i}\wedge d\bz_{i}\,,\label{ShortHands}
\end{eqnarray}
and we will make use of these forms in the calculations to follow.
In the large $\rho$ limit, $b(\rho)\sim
\rho\epsilon^{-4}\exp(-2\rho)\sim r^{-6}\log(r)$
and one finds that the potential $B_2$ varies logarithmically with
$r$, as in \cite{KT}.

\section{Supersymmetry Conditions}\label{SUSYembed}
\paragraph{}
Given the rich physics that has arisen from the study of D7-branes
in warped flux compactifications, it is interesting to find as many
different stable configurations of D7-branes in these backgrounds as
is possible. Because supersymmetry guarantees stability, and is
typically a simpler property to check than non-supersymmetric
stability, we will focus on a search for supersymmetric D7-branes in
warped throats with flux.  We begin by reviewing the supersymmetric
embedding conditions given in
\cite{MMMS,Open} for D7 branes wrapping a four-cycle in a generic Calabi-Yau
three-fold, and then in the next section we will apply these
conditions to specific examples of embeddings in the warped deformed
conifold.

Consider a CY three-fold $\cM_6$ with K\"ahler two-form $J$, a
background NS-NS 2-form potential $B_2$, and some number of
coincident D7-branes which fill out the four transverse Minkowski
space-time dimensions and extend in four directions on the compact
manifold. There is a vector field $A$ supported on the D7-brane
worldvolume, and its curvature $F_2 = dA$ combines with the
background $B_2$ to form a gauge invariant combination $\hcF$.
\begin{equation}
\hcF=\hat{B}_2+2\pi\alpha ' F_2\label{DefF}\,,
\end{equation}
where
we use $\hat{~}$ to denote the pull-back of bulk quantities on to
the D7-brane worldvolume.

In \cite{MMMS,Open}, it was found that the $\kappa$-symmetry
condition for D7 branes wrapping
a four-cycle $\Sigma_4$ in $\cM_6$ are as follows:
\begin{enumerate}
\item {\rm The four cycle $\Sigma_4$ must be holomorphic.  In other words,
the D7-brane locus may be written as a holomorphic equation in
the complex structure of the CY.}\label{cond1}
\item {\rm The generalized $\hcF$ field strength is a two-form of pure $(1,1)$ type, or}
\begin{equation}
\hcF^{2,0}=\hcF^{0,2}=0\,.\label{cond2}
\end{equation}
\item {\rm The K\"ahler form $J$ pulled back to $\Sigma_4$ and the
generalized field strength $\cF$ satisfy the equation:}
\begin{equation}
e^{-A}\hJ\wedge\hcF=\tan\theta\left(\frac{e^{2A}}{2}\hJ\wedge \hJ
-\frac{1}{2}\hcF\wedge\hcF\right)\label{cond3}\,.
\end{equation}
\end{enumerate}
In (\ref{cond3}) we have included the warp factor $e^A$ in
(\ref{10warpedmetric}) and the constant $\theta$ is specified by the
supersymmetry conditions for a given supergravity background
solution \cite{MMMS}.  These conditions are equivalent to the
requirement that the D7-brane configuration is a local energy
minimum, and in the fluxless case are equivalent to the condition
that $\Sigma_4$ is a minimal surface.

\paragraph{}
$\hcF$ must also satisfy a modified Bianchi identity:
\begin{equation}
d\hcF=\hat{H}_3\label{AnomalyFreeCond}\,,
\end{equation}
where $\hat{H}_3$ is the pull-back of $H_3=dB_2$.
Equation (\ref{AnomalyFreeCond}) arises from the definition of
$\hcF$ when the gauge field strength by itself is closed and
is necessary to guarantee cancellation of the global anomaly for
open fundamental strings ending on the D7 branes
\cite{FreedWitten}. Taken together, the condition (\ref{AnomalyFreeCond}) and the
supersymmetry conditions
are sufficient
conditions for the equations of motion from the D7 brane action to
be satisfied.

In the case of the warped deformed conifold, the parameter $\theta$
is zero and the condition (\ref{cond3}) simplifies to
\begin{equation}
\hJ\wedge\hcF=0\,,\label{cond3s}
\end{equation}
$\hcF$ satisfying (\ref{cond3s}) is said to be ``primitive.''  An
alternative and very useful interpretation
\cite{MMMS} for conditions (\ref{cond2}) and (\ref{cond3s}) is that
the generalized field strength $\hcF$ is anti-selfdual:
\begin{equation}
\hcF=-\ast_4 \hcF \label{selfdual}\,,
\end{equation}
where the Hodge star operation $\ast_4$ is taken with respect to the induced metric on the four cycle.

In terms of complex embedding coordinates $\{z_i\}$, $J$ and $B_2$
are manifestly $(1,1)$.
Therefore in the absence of the magnetic field $F_2$, $\cF$ is also
of the type $(1,1)$ satisfying the supersymmetry condition Eqn.
(\ref{cond2}).

\section{Supersymmetric D7 Embeddings in the Conifold}\label{SysSUSYD7}
\paragraph{}

In this section we shall examine several holomorphic D7-brane
embeddings in the singular conifold that have appeared in the
literature. In the absence of background H-flux, the only
requirement for a D7-brane to be supersymmetric is holomorphy of its
embedding equation.  Several such holomorphic embeddings have been
proposed, and
in the singular conifold limit, the $\kappa$-symmetry conditions are
trivially satisfied, and in some cases the supercharges have been
explicitly constructed \cite{ACR}. However, the supersymmetry
conditions are more complicated when H-flux is turned on, and our
goal in this section is to explore this situation in detail.

We will focus first on linear holomorphic embeddings in the interest
of simplicity.  There are two natural classes of linear embeddings,
given by either the description of the conifold in terms of the
$z_i$ coordinates or the $w_i$ coordinates.  The embedding $z_1=\mu$
was studied in detail in
\cite{Kuper} and
it satisfies the supersymmetry conditions (\ref{cond3s}) even on the
deformed conifold without needing any additional world volume flux.
Curiously, among the commonly studied classes of embeddings it
appears to be the only one that does not require a flux.

Another simple class of linear embeddings was introduced in
\cite{Ouyang} and is most naturally expressed in the $w_i$
coordinates as, for example, $w_1 = \mu$.  When nontrivial H-flux
is turned on, this embedding is not supersymmetric without also
turning on the world volume field strength $F_2$.  We will exhibit the
necessary flux, thus demonstrating that this class of D7-branes
can in fact be made supersymmetric.  To make the calculations
tractable we have taken the large-radius (KT) limit of the warped
deformed conifold, but it would be interesting to study these branes
in the full KS geometry as well.

In the following section we will refer to the relevant D7-branes as
being in either the ``$z$-embedding'' or the ``$w$-embedding'',
depending on the natural coordinates of the embedding equation.
 Having understood the simplest D7-brane embeddings, we will also see
that many other more complicated embeddings will require worldvolume
flux to be supersymmetric.

\subsection{$z$-Embedding}\label{SecKuperembedding}
\paragraph{}
We begin by considering the embedding for the deformed conifold
proposed in \cite{Kuper}
\begin{equation}
z_4=\mu\,,~~~~\mu\in{\mathbb{C}}\,.\label{Kuperembedding}
\end{equation}
This embedding breaks the $SO(4)$ isometry group of the deformed
conifold down to the  $SO(3)$ which rotates $\{z_1,z_2,z_3\}$. It
was shown to be supersymmetric in the deformed conifold for real
$\mu$ \cite{Kuper}.  We review the calculation here, generalizing as
well to complex $\mu$.
\paragraph{}
In order to check the supersymmetric condition (\ref{cond3}),
the pull back of the NS-NS B-field for this embedding can be obtained from
\begin{equation}
\hat{\eta}_1=\eta_1\vline_{z_4=\mu}=B_{1\bar{1}}dz_1\wedge d\bz_1+B_{1\bar{2}}dz_1\wedge d\bz_2+B_{2\bar{1}}dz_2\wedge d\bz_1+B_{2\bar{2}}dz_2\wedge d\bz_2\label{KuperBfield}\,,
\end{equation}
where the various components are given by:
\begin{eqnarray}
B_{1\bar{1}}&=&-|z_3|^{-2}{(z_1\bz_3-z_3\bz_1)(z_2\bar{\mu}-\bz_2\mu)}\nn\,,\\
B_{1\bar{2}}&=&(z_3\bar{\mu}-\bz_3\mu)+|z_3|^{-2}\left(z_1\bz_3(z_1\bar{\mu}-\bz_1\mu)+\bz_2 z_3(z_2\bar{\mu}-\bz_2\mu)\right)\nn\,,\\
B_{2\bar{1}}&=&-(z_3\bar{\mu}-\bz_3\mu)-|z_3|^{-2}\left(\bz_1 z_3(z_1\bar{\mu}-\bz_1\mu)+z_2 \bz_3(z_2\bar{\mu}-\bz_2\mu)\right)\nn\,,\\
B_{2\bar{2}}&=&|z_3|^{-2}{(z_2\bz_3-z_3\bz_2)(z_1\bar{\mu}-\bz_1\mu)}\nn\,,\label{KuperBfields}
\end{eqnarray}
and
\begin{equation}
z_3=\sqrt{(\epsilon^2-\mu^2)-(z_1^2+z_2^2)}\,.\label{Kuperz3}
\end{equation}
The components of the K\"ahler two-forms $J$ are given by:
\begin{eqnarray}
\hat{\eta}_5=\eta_5\vline_{(z_4=\mu)}&=& J'_{1\bar{1}}\,dz_1\wedge d\bz_1+J'_{1\bar{2}}\,dz_1\wedge d\bz_2+J'_{2\bar{1}}\,dz_2\wedge d\bz_1+J'_{2\bar{2}}\,dz_2\wedge d\bz_2\,,\label{DefKuperJp}\\
\hat{\eta}_4=\eta_4\vline_{(z_4=\mu)}&=& J''_{1\bar{1}}\,dz_1\wedge d\bz_1+J''_{1\bar{2}}\,dz_1\wedge d\bz_2+J''_{2\bar{1}}\,dz_2\wedge d\bz_1+J''_{2\bar{2}}\,dz_2\wedge d\bz_2\,.\label{DefKuperJpp}
\end{eqnarray}
Here the various functions are given by:
\begin{equation}
J'_{1\bar{1}}=1+\frac{|z_1|^2}{|z_3|^2}\,,~~~~J'_{2\bar{2}}=1+\frac{|z_2|^2}{|z_3|^2}\,,
~~~~J'_{1\bar{2}}=\frac{z_1\bz_2}{|z_3|^2}\,,~~~~J'_{2\bar{1}}=\frac{\bz_1 z_2}{|z_3|^2}\,.\label{KuperJps}
\end{equation}
\begin{equation}
J''_{1\bar{1}}=-\frac{(z_1\bz_3-z_3\bz_1)^2}{|z_3|^2}\,,~~~~J''_{2\bar{2}}=-\frac{(z_2\bz_3-z_3\bz_2)^2}{|z_3|^2}\,,
~~~~J''_{1\bar{2}}=J''_{2\bar{1}}=-\frac{(z_1\bz_3-z_3\bz_1)(z_2\bz_3-z_3\bz_2)}{|z_3|^2}\,.
\end{equation}

By either tedious or computer assisted algebra, one now obtains
\beq
\hat{\eta}_5\wedge\hat{\eta}_1 &=& 0\,, \label{Kupercond331} \\
\heta_4\wedge \heta_1 &=& 0\,.\label{Kupercond33}
\eeq
From equations (\ref{Kupercond331}) and (\ref{Kupercond33}), we can
confirm  that the SUSY condition (\ref{cond3}) is satisfied by the
holomorphic embedding (\ref{Kuperembedding}), as noticed in
\cite{Kuper}. Moreover as $\hat{\cF}=\hat{B}_2$, the
Bianchi identity (\ref{AnomalyFreeCond}) is also trivially
satisfied.

\subsection{$w$-Embedding}\label{Secwembedding}
\paragraph{}
Let us now consider the holomorphic embedding given in \cite{Ouyang}:
\begin{equation}
w_1=\frac{z_1+iz_2}{\sqrt{2}}=\mu\,,~~~~\mu\in{\mathbb{C}}\,.\label{OYembedding}
\end{equation}
This embedding breaks the $SO(4)\cong SU(2)\times SU(2)$ rotational symmetry
of the deformed conifold to the  $U(1)$ subgroup which rotates $w_3$
and $w_4$ by opposite phases.
By combining
(\ref{OYembedding}) with the defining equation of the deformed
conifold, we can also deduce that
\begin{equation}
w_2=\frac{z_1-iz_2}{\sqrt{2}}=\frac{\epsilon^2-(z_3^2+z_4^2)}{2\mu}\,,\label{OYw2}
\end{equation}
which leaves us only $\{z_3\,,z_4\}$ or $\{w_3\,,w_4\}$ and their
complex conjugates as the independent variables parameterizing the
four cycle wrapped by the D7 brane. In this case we shall first
demonstrate that supersymmetry is broken and then using numerical
methods construct the additional world volume flux which restores
it.
\paragraph{}
To check the supersymmetry condition (\ref{cond3s}),
the pull-back of the NS-NS two form field $\hat{B}_2=b(\rho)\heta_1$ can be obtained by calculating
\begin{equation}
\hat{\eta}_1=\eta_1\vline_{(w_1=\mu)}=B_{3\bar{3}}\,dz_3\wedge d\bz_3+B_{3\bar{4}}\,dz_3\wedge d\bz_4+B_{4\bar{3}}\,dz_4\wedge d\bz_3+B_{4\bar{4}}\,dz_4\wedge d\bz_4\,,
\end{equation}
where the various components are:
{\small
\begin{eqnarray}
B_{3\bar{3}}&=&i\left\{(z_4\bz_3-z_3\bz_4)\left(1+\frac{|z_3|^2}{|\mu|^2}\right)
+\frac{z_3 z_4}{\r2\mu}(\bz_1+i\bz_2)-\frac{\bz_3 \bz_4}{\r2\bar{\mu}}(z_1-iz_2)\right\}\label{OYB33}\,,\\
B_{3\bar{4}}&=&i\left\{(z_4\bz_3-z_3\bz_4)\left(\frac{z_3\bz_4}{|\mu|^2}\right)-i(z_1\bar{z}_2-\bar{z}_1 z_2)
+(|z_3|^2+|z_4|^2)-\frac{z_3^2}{\r2\mu}(\bz_1+i\bz_2)-\frac{\bz_4^2}{\r2\bar{\mu}}( z_1-iz_2)\right\}\nn\,,\label{OYB34}\\
\\
B_{4\bar{3}}&=&i\left\{(z_4\bz_3-z_3\bz_4)\left(\frac{z_4\bz_3}{|\mu|^2}\right)+i(z_1\bz_2-\bz_1 z_2)
-(|z_3|^2+|z_4|^2)+\frac{z_4^2}{\r2\mu}(\bz_1+i\bz_2)+\frac{\bz_3^2}{\r2\bar{\mu}}( z_1-iz_2)\right\}\nn\,,\label{OYB43}\\
\\
B_{4\bar{4}}&=&i\left\{(z_4\bz_3-z_3\bz_4)\left(1+\frac{|z_4|^2}{|\mu|^2}\right)
-\frac{z_3 z_4}{\r2\mu}(\bz_1+i\bz_2)+\frac{\bz_3 \bz_4}{\r2\bar{\mu}}( z_1-iz_2)\right\}\label{OYB44}\,.
\end{eqnarray}}
Notice that $\hat{\eta}_1$ is now only invariant under the $U(1)$
subgroup, and the invariance can be made manifest by changing into
the coordinates $\{w_3,w_4,\bar{w}_3,\bar{w}_4\}$.
\paragraph{}
The components for the pull-back of the K\"ahler form $\hat{J}=\cK'(\rho)\heta_5+\cK''(\rho)\heta_4$ can also be obtained by calculating:
\begin{eqnarray}
\hat{\eta}_5=\eta_5\vline_{(w_1=\mu)}&=& J'_{3\bar{3}}\,dz_3\wedge d\bz_3+J'_{3\bar{4}}\,dz_3\wedge d\bz_4+J'_{4\bar{3}}\,dz_4\wedge d\bz_3+J'_{4\bar{4}}\,dz_4\wedge d\bz_4\,,\label{DefOYJp}\\
\hat{\eta}_4=\eta_4\vline_{(w_1=\mu)}&=& J''_{3\bar{3}}\,dz_3\wedge d\bz_3+J''_{3\bar{4}}\,dz_3\wedge d\bz_4+J''_{4\bar{3}}\,dz_4\wedge d\bz_3+J''_{4\bar{4}}\,dz_4\wedge d\bz_4\,.\label{DefOYJpp}
\end{eqnarray}
Here the various functions are given by:
\begin{equation}
J'_{3\bar{3}}=1+\frac{|z_3|^2}{|\mu|^2}\,,~~~~J'_{4\bar{4}}=1+\frac{|z_4|^2}{|\mu|^2}\,,~~~~
J'_{3\bar{4}}=\frac{z_3\bz_4}{|\mu|^2}\,,~~~~J'_{4\bar{3}}=\frac{z_4\bz_3}{|\mu|^2}\,.\label{OYJp}
\end{equation}
\begin{eqnarray}
J''_{3\bar{3}}&=&S_3\bar{S}_3\,,~~~~J''_{3\bar{4}}=S_3\bar{S}_4\,,~~~~J''_{4\bar{3}}=S_4\bar{S}_3\,,~~~~
J''_{4\bar{4}}=S_4\bar{S}_4\,,\\
S_3&=&\bz_3-\frac{(\bar{\epsilon}^2-(\bz_3^2+\bz_4^2))z_3}{2|\mu|^2}\,,~~~
S_4=\bz_4-\frac{(\bar{\epsilon}^2-(\bz_3^2+\bz_4^2))z_4}{2|\mu|^2}\,.
\label{OYJpp}
\end{eqnarray}
Similarly, $\heta_4$ and $\heta_5$ are now only invariant under the
$U(1)$ isometry rotating the phase of $w_3/w_4$. Given the
expressions for $\hB_2$, and the pull-back of the K\"ahler form $J$
for the embedding (\ref{OYembedding}), we can now calculate the
wedge products explicitly and check if the supersymmetric condition
(\ref{cond3}) is satisfied. Let us calculate them in turn:
\begin{eqnarray}
\hat{\eta}_5\wedge \hat{\eta}_1
&=&i\frac{(z_3\bz_4-z_4\bz_3)}{|\mu|^2}\left(|z_1|^2+|z_2|^2+|z_3|^2+|z_4|^2\right)d\Omega\nn\\
&=&-\frac{|\epsilon|^4\sinh\rho\cosh\rho\sin\left(\frac{\theta_1+\theta_2}{2}\right)\sin\left(\frac{\theta_1-\theta_2}{2}\right)}{|\mu|^2}d\Omega
\label{OYpcond3}\,
\end{eqnarray}
with $d\Omega=dz_3\wedge dz_4\wedge d\bz_3\wedge d\bz_4$, and we
have used the explicit coordinates on the deformed conifold given in
Appendix \ref{DefConcoord}. Notice that the overall expression in
(\ref{OYpcond3}) is in fact real as
$i(z_3\bz_4-z_4\bz_3)=2{\rm{Im}}(z_4\bz_3)$. Similarly
\begin{eqnarray}
\hat{\eta}_4\wedge \hat{\eta}_1
&=&-\frac{i(z_3\bz_4-z_4\bz_3)}{|\mu|^2}\sum_{i\neq j=1}^{4}(z_i\bz_j-z_j\bz_i)^2 d\Omega\nn\\
&=&-\frac{|\epsilon|^6\sinh^3\rho\sin\left(\frac{\theta_1+\theta_2}{2}\right)\sin\left(\frac{\theta_1-\theta_2}{2}\right)}{|\mu|^2}d\Omega
\,.\label{OYppcond3}
\end{eqnarray}
Here we notice that the summation in (\ref{OYppcond3}) is in fact
positive definite, since each term in the summation (including the
negative sign) is nothing but $({\rm Im}(z_i\bz_j))^2$. Combining
(\ref{OYpcond3}) and (\ref{OYppcond3}), we can finally write down
the expression for $\hat{J}\wedge\hat{B}_2$ (including the scalar
functions):
\begin{equation}
\hat{J}\wedge\hat{B}_2=-\cK'(\rho)b(\rho)
\frac{|\epsilon|^4\sin\left(\frac{\theta_1+\theta_2}{2}\right)\sin\left(\frac{\theta_1-\theta_2}{2}\right)}{3|\mu|^2}
\sinh^2\rho \left(\frac{5+\cosh(2\rho)}{\sinh(2\rho)-2\rho}\right)d\Omega\,,\label{wJwedgeB}
\end{equation}
where the scalar functions $\cK'(\rho)$ and $b(\rho)$ are as given
in (\ref{DefKahPot1}) and (\ref{DefB}). The expression
(\ref{wJwedgeB}) is manifestly non-vanishing and we conclude that in
the absence of additional worldvolume flux $F_2$, supersymmetry is
broken.
\paragraph{}

\subsection{Restoring Supersymmetry with Worldvolume Flux}
\paragraph{}
Because $\hat{B}_2$ is not primitive, it is clear that D7-branes in
the $w$-embedding are not supersymmetric without the addition of
worldvolume flux.  In this section we will attempt to construct the
necessary flux.  After setting up the problem, we explore the
asymptotic behavior of the flux in sections
 \ref{bc1}-\ref{subleading}
before turning to a numerical solution in
section \ref{numeric}. The results are summarized in section
\ref{summary}.  Part of our
 solution agrees in appropriate asymptotic regimes with
 a previous proposal in \cite{Benini}, with some corrections which
 we describe in detail.

Our strategy is to first find a basis of 2-forms on the D7
worldvolume which are (1,1) and primitive. Then we take a linear
combination of these forms and impose the Bianchi identity, which
gives a system of partial differential equations for the
coefficients of the basis forms, and we proceed to solve this
system.

To make the calculations more tractable we will suppose the
parameter $\mu$ is much greater than the deformation parameter
$\epsilon$ so that we may work in the KT region of the deformed
conifold.  In this region, the background geometry is simply that of
the warped conifold with no deformation but with nontrivial
background fluxes.  The advantage of taking this limit lies in the
symmetries of the background.  In taking $w_1 = \mu$ on the deformed
conifold the isometries of $T^{1,1}$ are broken down to a single
U(1), but in the KT limit the isometry is $U(1)\times U(1)$.  In
addition to the $U(1)$ on the deformed conifold which rotates $w_3$
and $w_4$ by opposite phases, there is an extra $U(1)$ in the KT region
which acts as
$w_2 \rightarrow e^{2i\alpha}w_2, w_3\rightarrow e^{i\alpha}w_3, w_4
\rightarrow e^{i\alpha}w_4$.  The $U(1)^2$ isometry implies that
instead of having a four-variable problem, we can use the isometry
to eliminate the azimuthal angles, reducing our problem to a problem
of two spatial variables.

It is useful to express our embedding in terms of explicit
coordinates on the conifold.  Upon restricting to $w_1 = r^{3/2}
e^{i/2(\psi-\phi_1-\phi_2)}\sin\frac{\theta_1}{2}\sin\frac{\theta_2}{2}=\mu$,
we see that we may eliminate
\beq
r=\left(\frac{\mu}{\sin\frac{\theta_1}{2}\sin\frac{\theta_2}{2}}\right)^{2/3},\qquad
\psi =\phi_1+\phi_2,\label{wConstraints}
\eeq
and take $\mu$ to be real.  In terms of these angular coordinates,
large $r$ corresponds to $\theta_i
\rightarrow 0$ and when $\theta_1=\theta_2=\pi$ we are at the
minimal radius to which the D7-brane extends.

When pulled back to the locus $w_1=\mu$, the $\Omega_{ij}$ in
(\ref{Omega}) may be written as
\beq
\Omega_{11} &=& -2i\mu^2 \frac{dw_4 \wg d\bar{w_4}}{(|w_4|^2+\mu^2)^2} \\
\Omega_{22} &=& -2i\mu^2 \frac{dw_3 \wg d\bar{w_3}}{(|w_3|^2+\mu^2)^2}\\
\Omega_{12}+\Omega_{21} &=& -2i\mu^2\frac{w_3\bar{w_4}dw_4\wg d\bar{w_3}+w_4\bar{w_3}dw_3\wg
d\bar{w_4}}{|w_3||w_4|(|w_3|^2+\mu^2)(|w_4|^2+\mu^2)}
\eeq
To check primitivity, we need the Kahler form (up to a factor)
pulled back to the D7 brane worldvolume :
\beq
\hat{J} \propto Q_1 \Omega_{11} + Q_2
\Omega_{22}+\cot\frac{\theta_1}{2}\cot\frac{\theta_2}{2}
\left(\Omega_{12}+\Omega_{21}\right)
\eeq
where
\beq
Q_i = \frac32 + \cot^2 \frac{\theta_i}{2}.
\eeq
There are two simple anti-self-dual (1,1) forms which are relevant
to us:
\beq
X_1 &=& \Omega_{11} +\frac12 \tan \frac{\theta_1}{2}
\tan\frac{\theta_2}{2} Q_2 \left(\Omega_{21}+\Omega_{12}\right)\\
X_2 &=& \Omega_{22} +\frac12 \tan \frac{\theta_1}{2}
\tan\frac{\theta_2}{2} Q_1 \left(\Omega_{21}+\Omega_{12}\right)
\label{X1X2}
\eeq
There is also a third anti-self-dual (1,1) form, but it plays no
role in the later calculations:
\beq
X_3 = d\theta_1 \wg d\theta_2 -\sin\theta_1 \sin\theta_2 d\phi_1 \wg
d\phi_2.
\eeq
It is not hard to see that any component of $X_3$ in $\hcF$ will
either induce a violation of the Bianchi identity or will be
singular.

We can take the following linear combinations of $X_1$ and $X_2$ which have nice properties:
\beq
&& P = Q_1 X_1 -Q_2 X_2= Q_1\Omega_{11}-Q_2\Omega_{22}\,,\\
&& Q = X_1 -X_2\,.
\label{defPQ}
\eeq
The form $P$ is closed, $dP=0$, while $Q$ has a particularly simple
exterior derivative:
\beq
dQ = \frac{d\theta_1}{\sin\theta_1} \wg
\Omega_{22}-\frac{d\theta_2}{\sin\theta_2} \wg \Omega_{11}\,.
\label{dQ}
\eeq
We will use $P$ and $Q$ as basis two-forms for the worldvolume flux,
and search for an anti-self-dual form $\hcF = \alpha(\theta_1,\theta_2) P +
\beta(\theta_1,\theta_2) Q$ which satisfies the Bianchi identity.
It is enlightening to write the forms $P, Q$ explicitly in complex
coordinates:
\beq
\frac{P}{2i\mu^2} &=& \left(\frac32 + \frac{|w_3|^2}{\mu^2}\right)\frac{dw_3 \wg d\bar{w_3}}{(|w_3|^2+\mu^2)^2}
-\left(\frac32 + \frac{|w_4|^2}{\mu^2}\right)\frac{dw_4 \wg
d\bar{w_4}}{(|w_4|^2+\mu^2)^2}\label{Pwcoords}\\
\frac{Q}{2i\mu^2} &=&\frac{dw_3 \wg d\bar{w_3}}{(|w_3|^2+\mu^2)^2}-\frac{dw_4 \wg
d\bar{w_4}}{(|w_4|^2+\mu^2)^2} \nonumber \\
&&\qquad +\left(\frac{\mu^2}{|w_3|^2}-\frac{\mu^2}{|w_4|^2}\right)
\frac{w_3\bar{w_4}dw_4\wg d\bar{w_3}+w_4\bar{w_3}dw_3\wg
d\bar{w_4}}{(|w_3|^2+\mu^2)(|w_4|^2+\mu^2)}
\label{Qwcoords}
\eeq
By counting powers of $w_{3,4}$, we see that in the limit $w_3,w_4
\rightarrow \infty$, $Q$ is non-singular while $P$ is potentially
log-singular.  On the other hand, when $w_3$ or $w_4$ vanishes,
$P$ is non-singular while $Q$ can be log-singular.

If one prefers to parameterize the D7-brane worldvolume by the
angular coordinates $(\theta_1,\phi_1,\theta_2,\phi_2)$, the basis
forms may be written as
\beq
P&=& \left(\frac32 +\cot^2\frac{\theta_1}{2}\right) d\theta_1 \wedge
\sin\theta_1 d\phi_1- \left(\frac32 +\cot^2\frac{\theta_2}{2}\right) d\theta_2 \wedge
\sin\theta_2 d\phi_2\\
Q &=&\frac12 \left(\tan\frac{\theta_1}{2}\cot
\frac{\theta_2}{2}-\cot\frac{\theta_1}{2}\tan
\frac{\theta_2}{2}\right) \left(d\theta_2 \wedge
\sin\theta_1 d\phi_1+d\theta_1 \wedge
\sin\theta_2 d\phi_2\right)\cr
&&\qquad +d\theta_1 \wedge
\sin\theta_1 d\phi_1-  d\theta_2 \wedge
\sin\theta_2 d\phi_2.
\eeq

Let us now turn to the Bianchi identity which $\hcF$ must satisfy.
The B-field in the Klebanov-Tseytlin geometry is $B_2= \frac{3g_s
M\alpha'}{2} \log\frac{r}{r_0}\omega_2$, so that pulling back
$H_3=dB_2$ to the D7-brane worldvolume we have
\beq
d\hcF = \hat{H}_3 =-\frac{g_s
M\alpha'}{2}\left(\cot\frac{\theta_1}{2}d\theta_1 +
\cot\frac{\theta_2}{2} d\theta_2\right)\wg \omega_2
\label{bianchiangles}
\eeq
Plugging in the ansatz $\hcF = \alpha(\theta_1,\theta_2) P +
\beta(\theta_1,\theta_2) Q$, we obtain a first-order system of partial
differential equations for $\alpha(\theta_1,\theta_2)$ and
$\beta(\theta_1,\theta_2)$:
\beq
&&\partial_{\theta_1} \alpha = -\frac{S}{Q_2} \partial_{\theta_2}
\beta -\frac{1}{Q_2} \partial_{\theta_1} \beta + \frac{1}{Q_2 \sin
\theta_1} \beta
+ \frac{k}{Q_2} \cot \frac{\theta_1}{2}\,, \nonumber \\
&&\partial_{\theta_2} \alpha = \;\; \frac{S}{Q_1}
\partial_{\theta_1} \beta -\frac{1}{Q_1} \partial_{\theta_2} \beta +
\frac{1}{Q_1 \sin \theta_2} \beta + \frac{k}{Q_1} \cot
\frac{\theta_2}{2}.
\label{firstord}
\eeq
where we have
\beq
S=\frac{\cos\theta_2 - \cos\theta_1}{\sin\theta_1\sin\theta_2},
\qquad k = -\frac{g_s M\alpha'}{4}.
\eeq

This system of differential equations is challenging to solve but we
will show that numerical methods combined with some analytic tricks
will allow us to find a solution.  The system with two functions can
be converted to a second order partial differential equation for the
function $\beta(\theta_1,\theta_2)$ by differentiating and
eliminating $\partial_{\theta_1}\partial_{\theta_2} \alpha$.
Explicitly, the second order equation of interest takes the form
\beq
&&\Bigg[\partial_{\theta_1} \frac{S}{Q_1} \partial_{\theta_1}+
\partial_{\theta_2} \frac{S}{Q_2} \partial_{\theta_2}-
\partial_{\theta_1} \frac{1}{Q_1} \partial_{\theta_2}+\partial_{\theta_2} \frac{1}{Q_2}
\partial_{\theta_1}
+\frac{1}{Q_1\sin\theta_2}\partial_{\theta_1}-\frac{1}{Q_2\sin\theta_1}\partial_{\theta_2}
\nonumber\\
&& \qquad
\qquad+\frac{1}{\sin\theta_2}\partial_{\theta_1}\left(\frac{1}{Q_1}\right)
-\frac{1}{\sin\theta_1}\partial_{\theta_2}\left(\frac{1}{Q_2}\right)\Bigg]
\beta(\theta_1,\theta_2) \nonumber \\
&&\qquad \qquad =
-k\left(\cot\frac{\theta_2}{2}\partial_{\theta_1}\left(\frac{1}{Q_1}\right)
-\cot\frac{\theta_1}{2}\partial_{\theta_2}\left(\frac{1}{Q_2}\right)\right).
\label{B2ndord}
\eeq
Our main task in the following subsections will be to solve this
complicated equation.

Once we have solved for $\beta(\theta_1,\theta_2)$, we can then find
the second function $\alpha(\theta_1,\theta_2)$ by the following
procedure. From the system (\ref{firstord}), we construct the
quantity $(\partial_{\theta_1}^2 +
\partial_{\theta_2}^2)\alpha$.  This procedure results in a standard Poisson-type
 equation for $\alpha(\theta_1,\theta_2)$ with a complicated source term which
depends on the function $\beta(\theta_1,\theta_2)$.

\subsubsection{Boundary Conditions and Finiteness}
\label{bc1}

Before turning to the numerical analysis of the flux equations, we
need to identify the boundary conditions which we expect the
function $\beta(\theta_1,\theta_2)$ to satisfy (Recall that the
angular coordinates are defined over the range $(0,\pi)$.)  We claim
that there is a unique set of physical boundary conditions on the
supersymmetry-restoring flux.  For the function $\beta$, the
boundary conditions turn out to be compatible in a certain
large-radius limit with the proposal of Benini \cite{Benini}.

The primary physical constraint on $\beta$ is that we expect its
contribution to the  flux, $\beta Q$, to be finite.  One then finds
that the allowed boundary behavior of $\beta$ is tied to the
asymptotic properties of the two-form $Q$. There are three important
regimes to consider:
\begin{itemize}
\item {\bf $\theta_1=\pi$ or $\theta_2 = \pi$:}
When one or the other of the angles $\theta_i = \pi$, the only
physical choice of boundary conditions is Dirichlet:
\beq
\beta(\pi,\theta_2) = \beta(\theta_1,\pi) = 0.
\label{Dirichlet}
\eeq
The simplest argument for this comes from examining the basis
two-form $Q$ in (\ref{Qwcoords}).  For example, in the limit
$\theta_1 \rightarrow \pi$, we have $|w_4| \rightarrow 0$ so the
basis form $Q$ exhibits a logarithmic divergence.  To exclude this
singular flux, we must set $\beta=0$.
\item {\bf $\theta_2 \rightarrow 0$ with $\theta_1>0$, or vice
versa:}  In this limit, $|w_4|\rightarrow
\infty$ with $|w_3|$ finite, and one sees from (\ref{Qwcoords})
that $Q$ is $O(1)$. Thus the only constraint in this limit is
that $\beta$ is finite.
\item {\bf $\theta_1,\theta_2 \rightarrow 0$ simultaneously:}
In this limit, $Q$ actually vanishes as $1/|w|^2$, with
$|w_{3,4}|\rightarrow \infty$.  Then finiteness of the flux only
requires that $\beta$ grows no faster than $\frac{1}{\sin^4
\theta_i}$ as $\theta_i \rightarrow 0$.
\end{itemize}
To specify the large radius (small $\theta$) boundary conditions
precisely, we will now study the second and third limiting cases in
detail. The second case may be usefully regarded as a
``factorization'' limit in which the D7-brane worldvolume splits
into two branches.  Along one branch we have $\theta_1 \rightarrow
0$ and along the other, we have $\theta_2\rightarrow 0$.  The two
branches connect when both $\theta_1, \theta_2 \rightarrow 0$,
corresponding to the third case, which represents an
``interpolating'' limit between the two branches.

For the second function $\alpha$, the associated asymptotics may be
studied similarly.  The basis form $P$ is regular in the interior of
the D7-brane worldvolume, but has logarithmic singularities when
$\theta_1$ or $\theta_2$ vanishes.  This logarithm is physically
acceptable, as it is compatible with the log growth of the fluxes in
the KT solution.  Therefore when either angle vanishes, we require
that $\alpha$ grows no faster than $\log r$.  On the other hand, in
the interior of the brane, we expect $\alpha$ to be finite.

\subsubsection{Large Radius ``Factorization'' Limit}
\label{factorlim}

Let us take the limit $\theta_2 \rightarrow 0$, with $\theta_1$ held
fixed.  The various terms in (\ref{B2ndord}) have the behavior
\beq
&& S \rightarrow \tan\frac{\theta_1}{2} \frac{1}{\theta_2}\\
&& Q_2 \rightarrow \frac{4}{\theta_2^2}
\eeq
Multiplying the equation for $\beta$ by $\theta_2$ and taking the
limit as $\theta_2 \rightarrow 0$, we see that (\ref{B2ndord})
reduces to the following, discarding terms of order $\theta_2^2$:
\beq
&&\left[\partial_{\theta_1} \frac{1}{Q_1}
\tan\frac{\theta_1}{2}\partial_{\theta_1} +\frac{1}{4}\tan\frac{\theta_1}{2}
\theta_2\partial_{\theta_2}\theta_2\partial_{\theta_2}
-\partial_{\theta_1}\frac{1}{Q_1}
\theta_2\partial_{\theta_2}
+\frac{1}{Q_1}\partial_{\theta_1}
+\partial_{\theta_1}\left(\frac{1}{Q_1}\right)\right] \beta \cr &&
\qquad \qquad = -2k\partial_{\theta_1}\left(\frac{1}{Q_1}\right)
\label{betalim}
\eeq
Unfortunately this differential equation is not separable, but it
does have homogeneous scaling with $\theta_2$ -- all terms are of
zeroth order.

In the previous subsection we argued that in this limit $\beta$
should be finite. Thus the leading small $\theta_2$ behavior should
be for $\beta$ to be constant as a function of $\theta_2$.  The
equation (\ref{betalim}) simplifies to
\beq
\partial_{\theta_1} \frac{1}{Q_1}
\tan\frac{\theta_1}{2}\partial_{\theta_1}\beta
+\partial_{\theta_1}\left(\frac{\beta}{Q_1}\right)=-2k
\eeq
and the solution is
\beq
\beta = -2k + \frac{c_1}{2 \sin^2 \frac{\theta_1}{2}} \left(\cos\theta_1
 - 8 \log \sin\frac{\theta_1}{2}\right)+\frac{c_2}{2 \sin^2 \frac{\theta_1}{2}}
\label{firstsoln}
\eeq
with two integration constants which we need to determine.  We may
fix one constant by requiring consistency with the Dirichlet
boundary condition $\beta(\theta_1=\pi) = 0$, which is satisfied if
\beq
-2k-\frac{c_1}{2}+\frac{c_2}{2}=0.
\label{dconstraint}
\eeq
We will be able to obtain a second constraint after we have
performed the analysis in the next subsection.

Now we can substitute into the first order equations to solve for
$\alpha$. In the small $\theta_2$ limit these reduce to
\beq
\partial_{\theta_1} \alpha &=& -\frac14\tan\frac{\theta_1}{2} \theta_2
\partial_{\theta_2}
\beta  \label{smallt2firstord1}\\
\theta_2 \partial_{\theta_2} \alpha &=& \;\;
\frac{\tan\frac{\theta_1}{2}}{Q_1}
\partial_{\theta_1} \beta -\frac{1}{Q_1} \theta_2\partial_{\theta_2} \beta +
\frac{1}{Q_1} \beta + \frac{2k}{Q_1}.
\label{smallt2firstord2}
\eeq
The first equation reduces in our limit to $\partial_{\theta_1}
\alpha =0$ because $\beta$ is independent of $\theta_2$.  The second
equation with $\beta$ plugged in gives $\alpha \sim -2c_1\log
\theta_2$.

\subsubsection{Large Radius ``Interpolating'' Limit}
\label{largerad}

When $\theta_1$ and $\theta_2$ are simultaneously small, the
asymptotic analysis of section \ref{factorlim} breaks down.  For
example, there can be terms of the form $\frac{\theta_2}{\theta_1}$
which we discarded in the strict $\theta_2 \rightarrow 0$ limit but
are order one when the two angles are small simultaneously. This
immediately raises the question of whether it is possible for the
solutions along the two branches $\theta_1 =0$ and $\theta_2 =0$ to
be mutually consistent when both angles are small, so that the flux
smoothly interpolates between the two branches.  In the following
analysis we answer this question in the affirmative.

To proceed, we have found the following change of variables to be
useful:
\beq
u &=& - \log\left( \sin\frac{\theta_1}{2}\sin \frac{\theta_2}{2}\right)
\label{defu}\\
v &=& \log\left( \frac{\sin\frac{\theta_1}{2}}{\sin
\frac{\theta_2}{2}}\right)\label{defv}
\eeq
where $u$ is clearly related to the original radial coordinate $r$
as in (\ref{wConstraints})
The natural limit to take with both of the $\theta_i$ small is
simply
\beq
u \rightarrow \infty \, , \;\; v \; {\rm held\;  fixed}.
\label{interplim}
\eeq

Now we turn to the decoupled second order equation (\ref{B2ndord})
for $\beta$, in the $u$ and $v$ coordinates.  We may expand the
resulting equation in powers of $e^{-u}$ for large $u$, in the form
\beq
\left(\O_0 + e^{-u} \O_1 + e^{-2u} \O_2 + \ldots\right) \beta
= g_0 + e^{-u} g_1 + \ldots
\eeq
with homogeneous and inhomogeneous parts separated.  It is tedious
but straightforward to find (multiplying the equation obtained from
(\ref{B2ndord}) by an overall factor to reduce clutter)
\beq
\O_0 &=& \frac{\partial^2}{\partial v^2} + 2\coth v \frac{\partial}{\partial v}
+ 1\\
\O_1 &=& \frac{\cosh v}{4} \left(3 \frac{\partial^2}{\partial u^2}
-2\frac{\partial}{\partial u}-\frac{\partial^2}{\partial v^2}\right)
+\frac32 \sinh v \frac{\partial^2}{\partial u \partial v}\cr
&& \qquad -\frac12
\left(7\sinh v +\frac{1}{\sinh v}
\right)\frac{\partial}{\partial v} -2\cosh v\\
g_0 &=& -2k\\
g_1 &=& 4k \cosh v
\eeq
In principle one can now expand $\beta$ in powers of $e^{-u}$ and
construct a solution iteratively, order by order in $e^{-u}$.

At the zeroth order in the expansion, we encounter a small surprise:
in the large $u$ limit, the leading homogeneous part of the equation
for $\beta$ reduces to an ordinary differential equation,
\beq
\O_0
\beta_h =0.
\label{homeq}
\eeq
This equation is easily solved:
\beq
\beta_h = f_1(u) \frac{v}{\sinh v} + f_2(u) \frac{1}{\sinh v}
\label{leadingbetah}
\eeq
Demanding that $\beta$ is  non-singular at $v=0$ sets $f_2 =0$.

At leading order, $f_1(u)$ can be any function of $u$, but in the
full solution its form must be compatible with all the other
asymptotics we have identified.  This eliminates most possibilities
for $f_1$.  For example, a natural ansatz one might try is to set
$\beta= -2k + f_1(u)
\frac{v}{\sinh v} +O(e^{-u})$ with $f_1(u)$ of zeroth order in
$e^{-u}$, so that the leading homogeneous solution and the leading
particular solution are of the same order.  However, a little
manipulation shows that this ansatz must be wrong, because the
homogeneous part vanishes when $v \rightarrow \pm \infty$.  The
resulting function cannot possibly satisfy the Dirichlet boundary
condition (\ref{Dirichlet}).

What we need is a form for the function $f_1(u)$ so that the
homogeneous piece $\beta_h$ approaches a constant in the limit
$u\rightarrow \infty$, $v\rightarrow \pm \infty$ (which corresponds
to $(\theta_1,\theta_2) \rightarrow (0,\pi)$ or $(\pi,0)$.  There is
only one such function which suffices, up to subleading terms:
\beq
\beta_h = b e^{u} \left( \frac{1}{u} \frac{v}{\sinh v} \right)
+\ldots
\label{betah}
\eeq
Note that the growth of this $\beta_h$ for $u\rightarrow \infty$ is
sufficiently slow to be physically acceptable, as determined in
section
\ref{bc1}.  At the next order in $e^{-u}$, we find that the system
of equations is solved by
\beq
\beta = \beta_h -2k +O(1/u)
\label{betau}
\eeq
We will fix
the coefficient $b$ in the next subsection.

Let us now turn our attention to the second function $\alpha$. The
coupled first order equations for $\alpha$ and $\beta$ in the $u$
and $v$ variables, expanded to leading order in $e^{-u}$, are:
\beq
\partial_u \alpha &=& -e^{-u}\left(\cosh v \left(\partial_u \beta +
\beta + 2k\right) + 2 \sinh v \partial_v \beta \right) \\
\partial_v \alpha &=& -e^{-u} \left( \cosh v \partial_v \beta +
\sinh v (\beta + 2k) \right)
\eeq
To leading order at large $u$ (and including the first subleading
term in $1/u$), this is solved by
\beq
\alpha = \alpha_0 -2 b \log u - b \frac{ v \coth v}{u} +O(1/u^2)
\label{alphau}
\eeq
Because the basis form $P$ was closed by itself, $\alpha$ can always
be shifted by a constant $\alpha_0$.

One curious feature of the function $\beta$ which is clear from the
large-$u$ expansion is that the first subleading term for small
$\theta_2$ depends as $1/\log(\sin
\frac{\theta_2}{2})$; in radial coordinates this appears as
$1/\log(r)$ rather than as a power of $r$.  The function
$1/\log(\sin
\frac{\theta_2}{2})$ has the amusing property that although it
vanishes when $\theta_2 \rightarrow 0$, its first derivative with
respect to $\theta_2$ diverges.  We will make use of this subleading
behavior in the next section.

\subsubsection{Consistent Boundary Conditions and Subleading Asymptotics}
\label{subleading}

Having found asymptotic solutions in two different large radius
regimes, we now must check that these solutions are mutually
consistent.  Along the way, we will need to compute the
next-to-leading asymptotic behavior of the solutions to our PDEs.

The first additional consistency requirement will fix the ambiguity
we left in (\ref{firstsoln}).  There we had two integration
constants, $c_1$ and $c_2$, with one relation given by demanding
consistency with the ``small radius'' Dirichlet boundary condition.
To fix the second integration constant, we can take the simultaneous
limit $\theta_1
\rightarrow 0$ and $\theta_2 \rightarrow 0$ (recall that we have solved
for the exact $\theta_1$ dependence in the small $\theta_2$ limit)
and then compare with the results of the ``interpolating''
asymptotic; the results should agree.  For the limit of small
$\theta_2$, we can take (\ref{firstsoln}) with (\ref{dconstraint})
imposed to find
\beq
\beta \sim -(2k +c_1)\left(1-\frac{1}{\sin^2 \frac{\theta_1}{2}}\right)
-\frac{4c_1}{\sin^2 \frac{\theta_1}{2}}\log\sin
\frac{\theta_1}{2}
\label{b1}
\eeq
whereas for (\ref{betau}) written in the angular coordinates, we
find
\beq
\beta \sim \frac{2b}{\sin^2 \frac{\theta_1}{2}}\left(
1-2 \frac{\log\sin \frac{\theta_1}{2}}{\log\sin
\frac{\theta_2}{2}}\right) -2k
\label{b2}
\eeq
(In (\ref{b1}) and (\ref{b2}) the $\sim$ is meant to imply that
subleading terms of order $\left(\log\sin
\frac{\theta_2}{2}\right)^{-1}$have been dropped, although we have retained a term
proportional to $\frac{\log\sin \frac{\theta_1}{2}}{\log\sin
\frac{\theta_2}{2}}$.) Demanding that the terms proportional to a
constant times $\frac{1}{\sin^2
\frac{\theta_1}{2}}$ and $\frac{\log \sin\frac{\theta_1}{2}}{\sin^2
\frac{\theta_1}{2}}$ are in agreement sets \footnote{One slightly subtle point
is that in the factorization limit $\theta_2\rightarrow 0$, in
principle we should also collect contributions at all orders of
$e^{n(v-u)}$, from the point of view of the interpolating
asymptotic. However, the additional contributions which we drop are
not proportional to $\frac{1}{\sin^2
\frac{\theta_1}{2}}$ so they do not affect the matching of
 these particular terms.}
\beq
c_1 &=&0\\
b &=& k.
\eeq
Note that with $c_1=0$, we have
\beq
\beta= 2k \cot^2\frac{\theta_1}{2}
\label{leadingbeta}
\eeq
in the leading small $\theta_2$ limit.  This asymptotic for $\beta$
are identical to the proposal of \cite{Benini}.  The interested
reader will find a comparison of our notations in Appendix D.

To obtain a more detailed matching, and to see the leading
nontrivial behavior of $\alpha$, we must go on to the first
subleading order.  Expanding in the natural expansion parameter,
$1/\log \theta_2$, we write
\beq
\beta = -2k + \beta_0 + \frac{\beta_1}{\log\theta_2} + \frac{\beta_2}{\log^2\theta_2}+
\ldots
\eeq
The function $\beta_0$ was given in (\ref{leadingbeta}) and the
solution for $\beta_1$ has the same form:
\beq
\beta_1 = \frac{c_3}{2 \sin^2 \frac{\theta_1}{2}} \left(\cos\theta_1
 - 8 \log \sin\frac{\theta_1}{2}\right)+\frac{c_4}{2 \sin^2
 \frac{\theta_1}{2}}.
\label{secondsoln}
\eeq
As with the leading order, we set the two integration constants as
follows.  One constant is set by demanding consistency with the
small radius Dirichlet condition and the second is set by comparing
with the term proportional to $\frac{\log
\sin\frac{\theta_1}{2}}{\log
\sin\frac{\theta_2}{2}\sin^2
\frac{\theta_1}{2}}$ in (\ref{b2}) (the other terms cannot be matched
directly without working at higher order in the large-$u$ expansion,
because we have discarded $1/u$ corrections in $\beta_h$ which can
shift these terms.) These two constraints set
\beq
c_3 = c_4 = k.
\eeq
so
\beq
\beta_1=\frac{k}{ \sin^2 \frac{\theta_1}{2}}
\left(\cos^2\frac{\theta_1}{2}
 - 4 \log \sin\frac{\theta_1}{2}\right)
\eeq

Now, to find the leading behavior of $\alpha$, we plug into the the
equations (\ref{smallt2firstord1}), (\ref{smallt2firstord2}) and
solve. The answer is analytic but somewhat complicated:
\beq
\alpha= \alpha_0 -2k \log \left|\log {\theta_2\over 2}\right| +\frac{k}{\log^2 \theta_2}
\Big(-\log^2 \sin \frac{\theta_1}{2} + 2 \log\sin\frac{\theta_1}{2}
\log\cos\frac{\theta_1}{2}\qquad && \cr
 \qquad +\frac12 \log\sin\frac{\theta_1}{2} + \frac12 {\rm
Li}_2(\sin^2\frac{\theta_1}{2})\Big)&&
\label{alphasub}
\eeq
The leading $\log\log$ is consistent with the behavior in the
large-$u$ limit of (\ref{alphau}), but its appearance is perhaps a
bit surprising from the perspective of the ``factorization'' limit
in Section \ref{factorlim}.  In the factorization limit, at leading
order the $\log \log$ is not necessary to solve the field equations
(although it could have been included), but it is actually necessary
to include it to solve consistently at first subleading order.

\subsubsection{Numerical Solution}
\label{numeric}

In this subsection we construct a numerical solution for the
functions $\alpha$ and $\beta$ characterizing the D7-brane
worldvolume flux; the numerical computation gives strong evidence
that the D7-branes in the $w$-embedding are supersymmetric with the
addition of this flux.

In the preceding analysis, we identified a self-consistent set of
asymptotic solutions for the differential equation (\ref{B2ndord}).
The task that remains is for us to show that these asymptotics
correspond to a smooth solution over the entire worldvolume of the
D7-brane.

Under more fortunate circumstances, we would be able to invoke
existence and uniqueness theorems for solutions of this PDE, and the
task of showing that the D7-brane is supersymmetric would be
complete.  As it happens, the mathematical problem we must solve is
not of a standard type. The second-order PDE for $\beta$ is elliptic
through most of its domain, but it fails to be elliptic on the line
$\theta_1=\theta_2$, where the equation vanishes identically (recall
that the second order equation was obtained by differentiating and
subtracting the first order equations.)  We are not aware of general
theorems for this type of equation\footnote{Incidentally, the fact
the we obtained an ordinary differential equation rather than a PDE
in the limit of section \ref{largerad} was a sign of this
degeneracy.  Sometimes it is possible for such a degenerate equation
to be rendered elliptic by cleverly dividing by zero, but in this
case it is not clear to us that such a technique is applicable, as
it would make the asymptotic ODE (\ref{homeq}) singular.}.


In lieu of an abstract existence proof, we will attempt to
demonstrate the existence of a solution numerically, employing a
finite element method.  Strictly speaking, this method is only
guaranteed to converge for elliptic problems, but we will proceed
anyway under the assumption that near-ellipticity is enough.  The
danger that we face is that along the line $\theta_1 =
\theta_2$  where ellipticity fails, everything is a solution,
including spurious behavior, and the numerics can potentially be
quite unstable.


We have used the publicly available FreeFEM++ package
\cite{freefem} to perform the numerical calculations.  This software
uses the variational, or ``weak'' formulation of the finite-element
problem.  We use the default UMFPACK algorithm.  In the calculations
shown below, we have used a mesh with 2558 vertices, and we have
checked by varying the mesh that the solutions we obtain are not
mesh-dependent.

A trick that is useful in taming the spurious solutions of our
non-elliptic PDE is to impose Dirichlet boundary conditions whenever
possible. Fortunately the asymptotic analysis of the preceding
subsections allows us to do precisely this.  The boundary conditions
we have chosen are that $\beta$ vanishes when either angle $\theta_i
=\pi$ as in (\ref{Dirichlet}) and along the boundary where $\theta_i
\rightarrow 0$ we require that $\beta$ is given by the asymptotic
form in (\ref{betau}) with $b=k$ as determined by the analysis in
section \ref{subleading}.

Although the flux due to $\beta$  is finite, in our formulation the
function $\beta$ diverges as $\theta_1$ and $\theta_2$
simultaneously vanish, so we have solved instead for a rescaled
function $\gamma$ defined by
\beq
\gamma(\theta_1,\theta_2) = \left( 1-\cos^2{\theta_1 \over 2} \cos^2{\theta_2\over 2}
\right) \beta(\theta_1,\theta_2).
\eeq
which does not have this singular behavior.  We also impose  a small
cutoff, $\theta_i
> \theta_c = 10^{-7}$ to prevent problems with division by zero.  To be
specific, our boundary conditions for $\gamma$ take the form
\beq
&& \gamma(\theta_1,\pi) = \gamma(\pi,\theta_2) = 0\\
&& \gamma(\theta_1,\theta_{c}) = -2k +
\frac{2k}{\sin^2 \frac{\theta_1}{2}}
\frac{\log \sin\frac{\theta_{c}}{2}
-\log \sin\frac{\theta_1}{2}}{\log \sin\frac{\theta_{c}}{2} +\log
\sin\frac{\theta_1}{2}}\\
&& \gamma(\theta_c,\theta_{2}) = -2k +
\frac{2k}{\sin^2 \frac{\theta_2}{2}}
\frac{\log \sin\frac{\theta_{c}}{2}
-\log \sin\frac{\theta_2}{2}}{\log \sin\frac{\theta_{c}}{2} +\log
\sin\frac{\theta_2}{2}}
\eeq
This form was chosen to agree with the asymptotic solution
(\ref{leadingbeta}) in the limit $\theta_c\rightarrow 0$ and to be
compatible with the solution (\ref{betau}) in the ``interpolating''
limit.

\begin{figure}
\includegraphics[width=5in, height=3.5in]{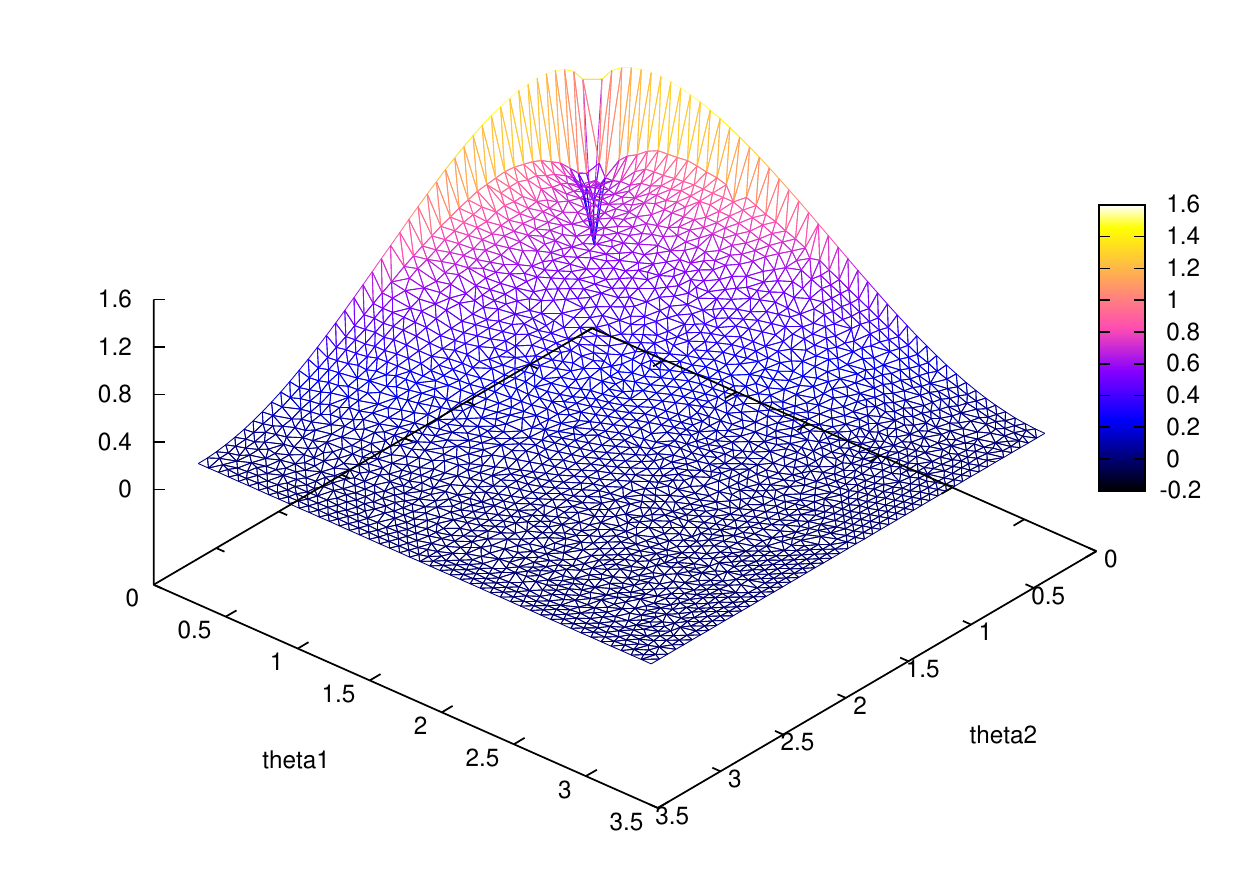}
\caption{\label{fig1} The $\gamma$ component of the flux, as found by a numerical calculation.
We have set $k=1$.}
\end{figure}

The result for $\gamma$ is shown in Figure \ref{fig1}.  We see that
the gradient of $\gamma$ appears to grow near the boundaries
$\theta_i
\rightarrow 0$, consistent with the subleading asymptotic dependence
$1/\log\theta_i$.  Near the origin, this effect becomes more
pronounced; we believe it reflects the behavior of the subleading
asymptotics in the $\theta_i$ coordinates and does not represent a
physical singularity.  The most important thing to take from this
diagram is that the solution exists and it appears to be free of
spurious behavior.

\begin{figure}
\includegraphics[width=5in, height=3.5in]{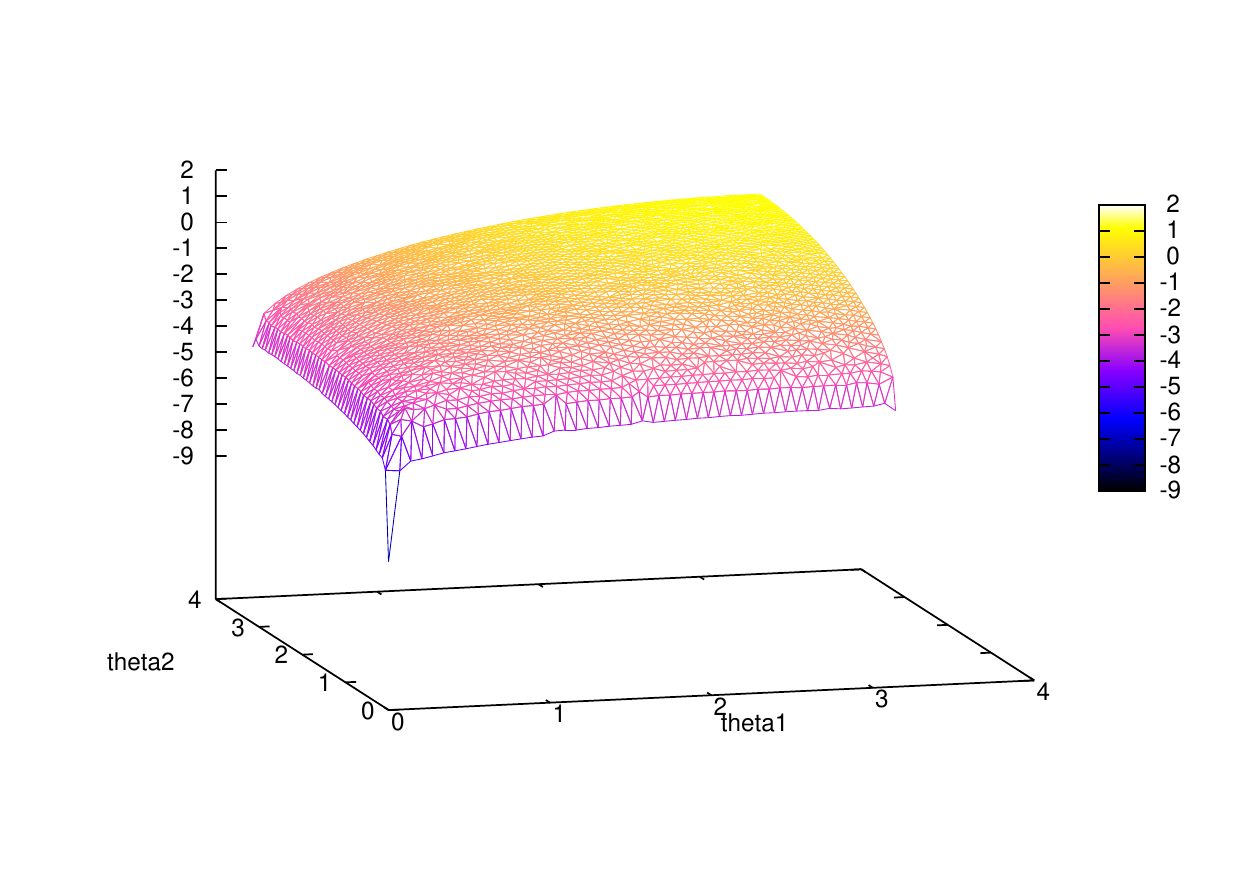}
\caption{\label{fig2} The $\alpha$ component of the flux, computed numerically in terms of the
angles $\theta_i$.  We have set $k=1$.}
\end{figure}

For the second function $\alpha$ we have implemented the procedure
of constructing a Poisson equation with a $\beta$-dependent source
term.
The boundary conditions on $\alpha$ are of generalized Neumann type
-- the normal derivatives of $\alpha$ must be compatible with the
first order system of equations (\ref{firstord}).  Strictly
speaking, these boundary conditions leave our problem slightly
ill-posed, but the ambiguity simply corresponds to shifting $\alpha$
by a constant, which was expected because the basis form $P$ was
closed.  Figure (\ref{fig2}) shows our result for $\alpha$. The
mildly singular behavior at the origin appears to be compatible with
the log $u$ dependence expected from (\ref{alphau}), and near the
corner $(\theta_1=\pi,\theta_2=\pi)$ it appears that $\alpha$
flattens out and approaches a constant.


\subsubsection{Summary of Results}
\label{summary}
Because the calculation of the supersymmetry-restoring flux was
quite complicated, let us pause to summarize the result.  The
self-dual (1,1) flux can be written as a linear combination of two
basis two-forms $P$ and $Q$ which are defined in (\ref{defPQ}).  We
have constructed a full solution numerically which appears to be
smooth everywhere in the interior of the brane worldvolume, and
which is presented graphically in Figures (\ref{fig1}) and
(\ref{fig2}).

The fact that our solution for the D7-brane flux is numerical is
somewhat unfortunate, as it makes the flux awkward to use in
calculations.  However, the numerical solution does have relatively
simple behavior at large and small radius, and we have given
analytic expressions for the large radius asymptotics.  The leading
asymptotics are consistent with the proposal of Benini
\cite{Benini}; we have extended his result by demonstrating that the
flux can be continued over the entire D7-brane worldvolume.

The small radius asymptotics ($\theta_1 \approx \pi$ and $\theta_2
\approx \pi$ simultaneously) are that $\beta=0$ and $\alpha$ is a
constant.

When one angle (say $\theta_2$) is taken to be small, the leading
behavior of $\beta$ is
\beq
\beta = 2k \cot^2 \frac{\theta_1}{2}
\eeq
which is consistent with the earlier proposal of \cite{Benini} (see
Appendix
\ref{appBenini}), and $\alpha$ has the leading behavior
\beq
\alpha = \alpha_0 -2k \log \left|\log {\theta_2\over 2}\right|.
\label{alphalimsmallt2}
\eeq
Here the $\log \log$ behavior gives a deviation from the result of
\cite{Benini}.

When both angles are small, the asymptotic analysis is somewhat
different. In terms of the coordinates defined by (\ref{defu}) and
(\ref{defv}), the leading asymptotic is given by
\beq
\beta = k e^{u} \left( \frac{1}{u} \frac{v}{\sinh v} \right) -2k
\label{betasum}
\eeq
where we have included the $-2k$ at the first subleading order in
$e^{-u}$, and the leading asymptotic for $\alpha$ is
\beq
\alpha = \alpha_0 -2 k \log u.
\label{alphasum}
\eeq
%


\subsection{Other Embeddings}
\paragraph{}
In this subsection we point out that several well-known examples of
D7-branes embeddings are not supersymmetric without the inclusion of
worldvolume gauge flux. It would be interesting, though presumably
quite difficult, to study whether a supersymmetry-restoring flux
exists in each case.

Interestingly, given a holomorphic embedding there is a simple
criterion that in many cases can immediately show that D7
worldvolume flux is necessary for supersymmetry.  When the D7-brane
is embedded holomorphically, it is possible to write its embedding
equation in terms of the $w_i$ coordinates: $g(w_i)=0$, say.  Then
if $g(w_i)$ is a polynomial, in the limit of large radius $r$ one
may approximate $g(w_i)$ by truncating to its terms of highest rank.
If the resulting homogeneous polynomial contains a factor of one of
the $w_i$ (up to an $SO(4)$ rotation), then in the large radius
limit the brane embedding admits a branch which locally looks like
the simple $w$-embedding, which as we have already seen requires a
nontrivial flux to be supersymmetric.

Let us now turn to some examples that have been studied in the
literature. The holomorphic embedding (\ref{Kuperembedding}) admits
a generalization that was proposed in
\cite{Kuper}
\begin{equation}
z_4=f\left(z_1^2+z_2^2\right)\,,\label{GKuperembedding}
\end{equation}
where $f(x)$ is an arbitrary function of $x=z_1^2+z_2^2.$ However,
we may also write $x=2(w_1w_2)$, so if $f(x)$ is a polynomial, then
in the large radius limit $f \sim (w_1 w_2)^p$ for some $p$, and it
may be approximated by $p$ copies of the $w_1=0$ and $w_2=0$
embeddings.  Based on the argument given above, we would expect this
embedding to be non-supersymmetric without a worldvolume flux, and
we have checked this explicitly in Appendix B.  A similar argument
applies to the Karch-Katz embedding \cite{Karch}:
\begin{equation}
w_1w_2=\frac{z_1^2+z_2^2}{2}=\mu^2\,,~~~~\mu\in{\mathbb{C}}\,,\label{KKembedding}
\end{equation}
and the interested reader will find details of the check in Appendix
C.
Finally, there is an interesting class of embeddings proposed by
Arean et al in \cite{ACR} given by
\begin{equation}
\prod^{4}_{i=1}w_i^{p_i}=\mu^{P}.\label{ACRembedding}
\end{equation}
This embedding admits a factorization in terms of the $w_i$ in the
large radius limit, so based on the examples we have studied so far
we expect that it will require a worldvolume flux to be
supersymmetric.

One might hope that, as in the case of the $w$-embedding, that the
necessary worldvolume fluxes might be constructible for more general
embeddings -- the fact that the flux we constructed vanishes for the
part of the D7-brane that dips into the throat is a promising sign.
The explicit calculations are daunting, however. A general
holomorphic embedding will break all the isometries of the conifold,
and the resulting system of equations one would need to solve to
find the flux can be a system of partial differential equations for
three functions of four variables.  Ideally one would hope for a
more abstract proof of existence for solutions, and we leave this as
an interesting but difficult open problem.

\section{Field Theory Remarks}\label{FieldTheory}
\paragraph{}
The D7-branes in the warped conifold geometry that we have studied
in the previous section have an interesting interpretation in the
dual gauge theory.  In this section we make some remarks about the
field theories for the various brane embeddings we have studied in
the earlier part of the paper (For a good review of cascading field
theories, see \cite{Strassler}.).

When no D7-branes are present, the dual gauge theory of the warped
deformed conifold is a non-conformal field theory with a product
gauge group  $SU(N+M) \times SU(N)$ and matter fields
$A_{1,2},B_{1,2}$ which transform in the bifundamental color
representations $({\bf N+M},{\bf \overline{N}})_c$ and $({\bf
\overline{N+M}},{\bf N})_c$.  The theory also has a superpotential
\beq
W=\lambda {\rm Tr}(A_i B_j A_k B_l) \epsilon^{ik} \epsilon^{jl}.
\label{UVKSpotential}
\eeq
When $M=0$, the gauge theory is actually superconformal \cite{KW}.  Moreover,
one can show that the  moduli space of the field theory corresponds
to the moduli space of $N$ D3 branes probing the conifold geometry.
To see this, take some number of D3-branes and move them away from
the tip of the conifold by giving expectation values to the fields
$A_i, B_j$.  Imposing the F-term and D-term equations, one discovers
that with the identifications
\begin{equation}
w_1=A_1 B_1,~~w_2=A_2 B_2,~~w_3=A_1 B_2,~~w_4=A_2 B_1\,.\label{Dictionary}
\end{equation}
we recover precisely the defining equation of the conifold
(\ref{conifw}).

When $M\neq 0$, conformal invariance is broken and the theory
undergoes renormalization group flow \cite{KS,KT}. Along the RG
flow, the two gauge group factors take turns being strongly coupled;
when one of them goes to strong coupling, one can perform a Seiberg
duality which results in a weakly coupled description. Each Seiberg
duality causes the gauge group to change from $SU(N+M)
\times SU(N)$ to $SU(N) \times SU(N-M)$ \cite{SeiDual}. Ultimately, in the far IR, this ``cascade" must end, as the gauge group ranks cannot be
negative. At that point, the theory undergoes confinement, and has a
dynamically generated scale which is related to the deformation
parameter of the conifold \cite{KS}.

When we add D7-branes in the gravity theory, the gauge theory is
correspondingly modified by the addition of a number of matter fields
charged as fundamentals of the gauge group. The introduction of
additional matter should change the duality cascade pattern, and it
was argued in
\cite{Ouyang} that the extra matter causes the duality
cascade to slow down as the RG scale decreases.


\subsection{Cascade Pattern for $z$ embedding and its Generalizations}\label{PatKupembed}
\paragraph{}
The analogue of the KS duality cascade when we include probe branes
embedded by $z_1=\mu$, is actually quite simple, and is manifestly
self-similar, as was pointed out in
\cite{Flavback}. Although the discussion in this subsection will
contain nothing new for experts, for the sake of explicitness, we
review here the cascade analysis.

Because the worldvolume flux can vanish for the supersymmetric
$z$-embedding, it is reasonable to guess that the appropriate field
theory description for the $z$-embedding in the KS background
differs from that of the singular conifold theory only by changing
the ranks of the gauge groups from $SU(N)\times SU(N)$ to
$SU(N+M)\times SU(N)$.  The superpotential for the case of one such
D7 brane (which one can motivate based on RG flow from a related
orbifold theory
\cite{Ouyang}) is
\begin{equation}
W_{z_1=\mu}= \lambda_1 q(A_\alpha B_\alpha-\mu)\tq+\lambda_2\frac{(q\tq)^2}{2}\,.\label{WfKuper}
\end{equation}
the additional quarks $q$ and $\tq$ transform in the $\bf{(N+M, 1)}$
and $\bf{(\overline{N+M},1)}$ representations of the $SU(N+M)\times
SU(N)$ gauge group (of course we can also choose $q$ and $\tq$ to be
charged under the $SU(N)$ gauge group instead.) One check of this
superpotential is that if we probe the theory with a D3-brane, then
the quarks, which one can think of as D3-D7 strings, become massless
precisely when the probe D3-brane intersects with the D7. The
superpotential has an obvious generalization to the case of $K$
coincident D7-branes. Let us consider the case where the additional
quark $q$ transforms in the ${\bf(N+M,1)}$ representation, so that
the quark is coupled to the $SU(N+M)$ factor which flows to strong
coupling. As in the case of the regular KS cascade, there are
mesonic operators $M_{\alpha\beta}= A_{\alpha}B_{\beta}$, and
because of the added quarks there are also operators which we label
as $C,D$ and $N$:
\begin{equation}
\left(C_\alpha\right)^m_a=\left(A_\alpha\right)^m_i\tq^i_a\,,~~~~
\left(D_\alpha\right)^a_m=q_i^a\left(B_\alpha\right)^i_m\,,~~~~
N^a_b=q_i^a\tq^i_b\,,~~i=1\dots N+M,~~a\,,b=1\,,\dots\,,K\,.\label{KuperGaugeInvs}
\end{equation}
These additional $SU(N+M)$-invariant quantities and along with the
original mesons $M_{\alpha\beta}$ allow us to rewrite the total
modified superpotential for the $z$ embedding into:
\begin{equation}
W_{Total}=h{\rm Tr}\left(M_{\alpha\beta}M_{\gamma\delta}\right)\epsilon^{\alpha\gamma}\epsilon^{\beta\delta}
+ \lambda_1{\rm Tr}(D_\alpha C_\alpha-\mu N)+\lambda_2\frac{N^2}{2}\label{WfKuper2}\,.
\end{equation}
Now, as the $SU(N+M)$ factor flows to strong coupling, we see that
we have $N_c=N+M$ and $N_f=2N+K$, so we are in the conformal window
and may perform a Seiberg duality.  Under this duality, the weakly
coupled theory with gauge group $SU(N-(M-K))\times SU(N)$, and
$2N+K$ flavors coupled to the $SU(N-(M-K))$ factor. Notice that in
contrast with the original KS theory, where the difference between
the ranks of two gauge groups remain unchanged throughout the
duality cascade, here the difference between the ranks reduces by
$K$ at each step of the cascade. The resulting dual superpotential
now receives the following additional contributions:
\begin{equation}
\frac{1}{\Xi}\left[{\rm Tr}(M_{\alpha\beta}a^{\alpha}b^{\beta})+{\rm Tr}(C_\alpha q'b^\alpha)+{\rm Tr}(D_\alpha a^\alpha \tq')+{\rm Tr}(Nq'\tq')\right]\,.\label{DWfAll}
\end{equation}
Here $a_{\alpha}$ and $b_{\alpha}$ are the dual fields for
$A_{\alpha}$ and $B_{\alpha}$, under the $SU(N-(M-K))\times SU(N)$
gauge group, they transform as ${(\bf N-(M-K),\overline{N})}$ and
${\bf(\overline{N-(M-K)},N)}$ representations respectively; whereas
the $K$ $q'$ and $\tq'$s, the dual quarks and anti-quarks for $q$
and $\tq$ transforms in the ${\bf(N-(M-K),1)}$ and
${\bf(\overline{N-(M-K)},1)}$ representations.
\paragraph{}
Now, we may integrate out the $SU(N-(M-K))$ singlets.  First, note
that one obtains the following F-term equations:
\begin{eqnarray}
M_{\alpha\beta}&:&2h\epsilon^{\alpha\gamma}\epsilon^{\beta\delta}M_{\gamma\delta}+\frac{1}{\xi}a^{\alpha}b^{\beta}=0\,,\label{eqnMKuper}\\
N&:& \lambda_2 N-\mu+\frac{1}{\xi}q'\tq'=0\,,\label{eqnNKuper}\\
C_{\alpha}&:& \lambda_1 D_{\alpha}+\frac{1}{\xi}q'b_{\alpha}=0\,,\label{eqnCKuper}\\
D_{\alpha}&:& \lambda_1 C_{\alpha}+\frac{1}{\xi}a_{\alpha} \tq'=0\label{eqnDKuper}\,.
\end{eqnarray}
Substituting (\ref{eqnMKuper})-(\ref{eqnDKuper}) into
(\ref{WfKuper2}) and (\ref{DWfAll}), we obtain the superpotential
for the weakly coupled $SU(N-(M-K))\times SU(N)$ theory:
\begin{equation}
W_{\rm Dual}=-\frac{1}{\xi^2}\left(\frac{1}{4h}{\rm Tr}(a_\alpha b_\beta a_\gamma b_\delta)\epsilon^{\alpha\gamma}\epsilon^{\beta\delta}+\frac{1}{\lambda_1}q'\left(b_\alpha a_\alpha-\frac{\xi\mu}{\lambda_2}\right)\tq'+\frac{1}{2\lambda_2}(q'\tq')^2\right)\label{DWfKuper}\,.
\end{equation}
This superpotential (\ref{DWfKuper}) has exactly the same functional
form (up to field re-definitions and extra constants) as the
original superpotential (\ref{UVKSpotential}) and (\ref{WfKuper}),
except that the quarks are charged under the gauge group which flows
to weak coupling.
\paragraph{}
Now let us note that if the additional quarks $q,\tq$ are coupled to
the gauge group that flows to weak coupling, the analysis proceeds
essentially as before, except that it is simpler because the quarks
$q$ and $\tq$ do not bind to form mesons. After performing a Seiberg
duality on the strongly coupled gauge group, the two gauge groups
switch roles again and the quarks $q,\tq$ will again be coupled to
the gauge group which flows to strong coupling.  Thus after two
steps of the cascade, the theory is fully self-similar.

\subsection{The vacua for Kuperstein-like Embeddings}\label{Kupvacua}
\paragraph{}
As was shown in \cite{Kuper} and reviewed in Section 4.1, the simple
$z$-embedding is supersymmetric for all $\mu$ on the full deformed
conifold.  Therefore it is interesting to study the case of a
D7-brane embedded by $z_1=\mu$ in the extreme infrared ($\mu \sim
\epsilon$), where the conifold becomes deformed.  In this case, the
quarks due to the D7-brane have masses of the order of the
confinement scale and can modify the IR dynamics of the theory.

Suppose the number of units of 3-form flux, after having
decreased from the D7 backreaction, is $M^*$\footnote{The explicit value $M^*$ can be deduced from the pattern of the flavor cascade described earlier to be $M^*=\sqrt{(M-K/2)^2-2NK}-K/2$.}.
To investigate the
moduli space of the theory, we use the standard trick of probing the
theory by adding a single mobile D3-brane.  The gauge group on the
D3 brane probe is $SU(M^*+1)\times U(1)$, and we have matter fields
$a_{\alpha}$ and $b_{\alpha}$ transforming respectively in
${\bf(M^*+1,1)}$ and ${\bf(1,M^*+1)}$ representations, moreover we
also have an additional quark $q'$ coming from 3-7 string transform
in ${\bf(M^*+1,1)}$. By self-similarity, the UV superpotential on
the probe here is essentially
identical to (\ref{DWfKuper}), but with the fields and the
parameters modified accordingly. At low energy, the UV
superpotential can be written again in terms of the $SU(M^*+1)$
invariant quantities and it also
acquires a non-perturbative Affleck-Dine-Seiberg superpotential
\cite{ADS}:
\begin{equation}
W_{\rm ADS}
=(M^*-2)\left(\frac{{\tilde{\Lambda}}^{3M^*}}{{\det} \Omega}\right)^{\frac{1}{M^*-2}}\label{KuperADS}\,.
\end{equation}
Here the determinant of the $3\times 3$ meson matrix
${\rm{det}}\Omega$ is given by
\begin{equation}
{\det}\Omega=\frac{N}{2}\epsilon^{\alpha\gamma}\epsilon^{\beta\delta}M_{\alpha\beta}M_{\gamma\delta}
-\epsilon^{\alpha\gamma}\epsilon^{\beta\delta}C_{\alpha}D_{\beta}M_{\gamma\delta}\,,\label{DetOmega}
\end{equation}
whereas the various gauge invariant quantities are defined similarly
as in (\ref{KuperGaugeInvs}). From (\ref{KuperADS}) and
(\ref{DetOmega}) we can deduce the $F$-term equations:
\begin{eqnarray}
M_{\alpha\beta}&:& \epsilon^{\alpha\gamma}\epsilon^{\beta\delta}(hM_{\alpha\beta}-S(\Omega)(NM_{\alpha\beta}-C_\alpha D_\beta))=0\,,\label{IRKuperFterm1}\\
N&:& \lambda_2 N+\frac{1}{2}S(\Omega)\epsilon^{\alpha\gamma}\epsilon^{\beta\delta}M_{\alpha\beta}M_{\gamma\delta}=\lambda_1\mu\,,\label{IRKuperFterm2}\\
C_\alpha&:& \lambda_1\epsilon^{\alpha\beta}D_{\beta}=S(\Omega)\epsilon^{\alpha\gamma}\epsilon^{\beta\delta}D_{\beta}M_{\gamma\delta}\,,\label{IRKuperFterm3}\\
D_\alpha&:& \lambda_1\epsilon^{\alpha\beta}C_{\alpha}=S(\Omega)\epsilon^{\alpha\gamma}\epsilon^{\beta\delta}C_{\alpha}M_{\gamma\delta}\,.\label{IRKuperFterm4}
\end{eqnarray}
with
\begin{equation}
S(\Omega)=\left(\frac{\tilde{\Lambda}^{3M^*}}{\det \Omega}\right)^{\frac{1}{M^*-2}}\frac{1}{\det\Omega}\label{DefSOmega}\,.
\end{equation}
One class of simple solutions to these F-term equations is given by
setting $C_\alpha=D_\alpha=0$ and we have
\begin{eqnarray}
&&N=\frac{1}{2\lambda_2}\left(\lambda_1\mu\pm\sqrt{(\lambda_1\mu)^2-4\lambda_2 h\det M_{\alpha\beta}}\right)\,,\nn\\
&&h\det M_{\alpha\beta}=\left(\frac{\tilde{\Lambda}^{3M^*}}{N\det M_{\alpha\beta}}\right)^{\frac{1}{M^*-2}}\,.
\end{eqnarray}
The deformed conifold branch naturally appears when we consider the
limit $(\lambda_1\mu)^2\ll 4\lambda_2 h\det M_{\alpha\beta}$, which
yields:
\begin{equation}
\det M_{\alpha\beta}\approx \left(\frac{\lambda_2\tilde{\Lambda}^{6M^*}}{h^{2M^*-3}}\right)^{\frac{1}{2M^*-1}}\label{KuperDefConifold}\,.
\end{equation}
Another extreme limit which can be taken is that
$(\lambda_1\mu)^2\gg 4\lambda_2 h\det M_{\alpha\beta}$, and again
simple algebra shows that we recover the deformed conifold. These limits
precisely correspond to moving the probe D3-brane away from the
flavor D7 branes, and locally it detects the deformed conifold geometry.
\paragraph{}
It would be very interesting to understand the other regions of the
moduli space especially with $C_{\alpha} ,D_{\beta} \neq 0$.

\subsection{$w$-embedding and its Generalizations}\label{Patwembed}
\paragraph{}
Now let us move to the cascade analysis for $w$-embedding and its
generalizations, such as the ACR embeddings \cite{ACR}. We shall
begin by relating
the superpotential for various ACR-embeddings
to the one for the simplest $w$-embedding (\ref{OYembedding}); we
then on the cascade pattern for the simplest $w$-embedding, as the
more complicated cases can be analyzed similarly.
\paragraph{}
Recall that for the $w$-embedding (\ref{OYembedding}), an
appropriate additional term to the KS superpotential
(\ref{UVKSpotential}) was given in \cite{Ouyang}:
\begin{equation}
W_{w_1=\mu}=\lambda q(A_1 B_1-\mu)\tq\label{WfOuyang}\,,
\end{equation}
where $q$ can transforms in the representation ${\bf(N+M,1)}$ and
$\tilde{q}$ transforms in the ${\bf(\overline{N+M},1)}$
representation of the gauge group. This superpotential can be
generalized to many other interesting cases.  Take for example the
Karch-Katz embedding
\cite{Karch}, $w_1 w_2 = \mu$. In the large radius limit, we may
approximate this embedding equation by $w_1w_2=0$, which factorizes
into two branches $w_1=0$ and $w_2=0$. On the field theory side,
this naturally corresponds to introducing two sets of additional
quarks $\{q_1\}$ and $\{q_2\}$ with the superpotential
\begin{equation}
\kappa_1 q_1(A_1B_1)\tq_1+\kappa_2 q_2(A_2B_2)\tq_2\,.\label{WfKK}
\end{equation}
At smaller radius, however, the two branches meet and at least one
set of quarks will be massive. We can describe this with
superpotential terms of the form
\begin{equation}
\gamma_1 q_2\tq_1+\gamma_2 q_1\tq_2~~{\rm with}~~\gamma_1\gg\gamma_2\,.\label{Qbilinear}
\end{equation}
Taking the larger of the two masses and integrating out the
associated quarks, we obtain a superpotential of the form
$q[(A_1B_1)(A_2B_2)+ \gamma_2]\tq$, as desired.  Of course if we
continue further and integrate out quarks using the $\gamma_2$ mass
term we will be left with a trivial superpotential.

Now, for the more general ACR embedding, we can implement a
generalization of the procedure we have described for the Karch-Katz
embedding.  Again, the ACR equation can be factorized into different
branches in the vanishing $\mu$ limit, each of which is given by
$w_i=0$ for some $i$, motivating superpotential terms of the form
\begin{equation}
\sum_{n_{1}=1}^{p_1}\kappa_{n_1}q_{n_1}(A_1B_1)\tq_{n_1}
+\sum_{n_{2}=1}^{p_2}\kappa_{n_2}q_{n_2}(A_2B_2)\tq_{n_2}
+\sum_{n_{3}=1}^{p_3}\kappa_{n_3}q_{n_3}(A_1B_2)\tq_{n_3}
+\sum_{n_{4}=1}^{p_4}\kappa_{n_4}q_{n_4}(A_2B_1)\tq_{n_4}\,,\label{WfACR2}
\end{equation}
which are simply linear combinations of the simplest $w$-embedding.
By adding appropriate mass terms and integrating out quarks as they
become massive, it is not hard to see that we can flow to a
superpotential of the form
\begin{equation}
W_{ACR}=\lambda q\left[(A_1B_1)^{p_1}(A_2B_2)^{p_2}(A_1B_2)^{p_3}(A_2B_1)^{p_4}-\mu^P\right]\tq
\,,\label{WfACR}
\end{equation}
which is related to the ACR embedding equation in a direct way.

\paragraph{}
Now, in the case with no three-form flux on the conifold, the D7
branes in the $w$-embedding factorizes further into two branches,
using the relation (\ref{Dictionary}) $w_1 = A_1 B_1$.
A superpotential reflecting this factorization property was proposed
in \cite{Ouyang} and studied in \cite{Benini}:
\begin{equation}
W_{flav}=g_1 QA_1\tq+g_2 qB_1 \tilde{Q}\label{UVwembpotential}\,.
\end{equation}
By introducing the quark mass terms $m_1 q\tq+m_2
Q\tilde{Q}\,,~m_2\gg m_1$ the familiar superpotential term
(\ref{WfOuyang}) (up to appropriate redefinition) can be recovered
by introducing the quark mass terms $m_1 q\tq+m_2 Q\tilde{Q}$ and
integrating out appropriately.
The cascade
pattern for this type of superpotential has been studied in
\cite{Benini}, with the result that the self-similarity under Seiberg
duality transformation no longer presents in this case and a
sequence of  additional terms involving singlet states and quarks is
generated. Explicitly the resulting superpotential after a step of
cascade is given by:
\begin{equation}
\lambda_1{\rm Tr}(a_\alpha b_\beta a_\gamma b_\delta)\epsilon^{\alpha\gamma}\epsilon^{\beta\delta}+ \lambda_2(\bar{P} a_2 q'+\tq' b_2 P+\Sigma_0 q'\tq')\label{UVwembpotential1}\,.
\end{equation}
Here $\lambda_{1,2}$ are combinations of coupling constants and
dynamical scale; $\Sigma_0$ is an additional gauge singlet field.
In the proposal of \cite{Benini}, the cascade is almost
self-similar; at each step the term with the singlet $\Sigma$
acquires additional $A,B$ fields and (at least naively) becomes more
and more irrelevant.  This unusual cascade deserves further study;
perhaps the worldvolume flux we constructed in section 4.3 may shed
some light on the nature of the field theory.

\section{Discussion}\label{Conclusion}
\paragraph{}

We conclude with a few comments on extensions and possible
applications of our results. Having seen that worldvolume fluxes can
restore supersymmetry, it would of course be interesting to study
fluxes for embeddings other than the simple linear $w$-embedding,
and to extend our calculations to the deformed conifold. As we have
already remarked, this is in general a daunting mathematical
exercise, but perhaps it will be possible to prove an existence
theorem of some kind for the necessary fluxes, possibly based on the
topology of the embedding. Understanding the worldvolume flux may
also help in extending the calculations of D7-brane backreaction
along the lines of \cite{Flavback,Bigazzi:2008zt}.  It would also be
interesting to study D7-branes in warped cones other than the
conifold.

In the gauge/gravity duality, these D7-branes add flavor to the dual
KS gauge theory, with corresponding prospects for understanding
mesons and baryons in a confining theory.  It might be interesting
to revisit the calculations of the meson spectrum, as for example in
\cite{Levi:2005hh}, with worldvolume fluxes turned on.  In
particular, it would be interesting to understand better the
physical meaning of the closed (1,1) form $P$ which we identified in
Section 4.3, which one can turn on without breaking any
supersymmetry; it should correspond to deformation of the field
theory by some operator. A clearer understanding of these
supersymmetry-restoring fluxes may also be important in clarifying
the nature of the duality cascade with probe $w$-D7-branes.

There are also potential applications to inflationary model
building. The model constructed in \cite{DelicateU} used a very
special brane embedding to generate the inflationary potential, but
with several different supersymmetric brane embeddings, one can
imagine combining D7-branes in various ways to obtain many other
inflationary models. It would also be very interesting to understand
if the presence of extra flux can modify the explicit expression for
the gaugino condensate given in
\cite{Baumann:2006th}. It was argued in
\cite{DelicateU} that at the leading order expansion of D7 DBI
action, the additional supersymmetric flux should not modify the
results obtained in \cite{Baumann:2006th}. Combining the severe
constraints of holomorphy of the superpotential and the global
symmetries preserved by the embedding,
one is tempted to conclude that the potential due to the gaugino
condensate in the deformed conifold should be quite similar to that
of the singular conifold. Having a supersymmetric D7 configuration
including the necessary world volume flux should allow us check such
a statement explicitly, following the calculations in
\cite{Baumann:2006th}.

Also, it would also be interesting to consider small deviations from
the extremal limit, and allow supersymmetry to be softly broken by a
D-term potential whose magnitude is set by $\int_{\Sigma_4} J\wedge
\cF$ where $\Sigma_4$ is the four cycle which D7 brane wraps on in
the deformed conifold \cite{BKQ} (see also \cite{Dasgupta:2008hw}
for an interesting construction in the resolved conifold.) Combined
with the fact that $\Sigma_4$ is generally warped, this may allow us
to consider a variety of interesting scenarios for supersymmetry
breaking.

\section*{Acknowledgments}
It is a pleasure to thank Francesco Benini, Anatoly Dymarsky,
Mikhail Feldman, Fernando Marchesano, Bret Underwood, and Angel
Uranga for helpful discussions. We also thank an anonymous referee
for very useful criticisms.  HYC would like to thank KITP at UCSB,
the Sixth Simons Workshop at SUNY Stony Brook and Institute for
Advanced Study, Princeton for their hospitality while part of the
work was being carried out. This work was supported in part by NSF
CAREER Award No. PHY-0348093, DOE grant DE-FG-02-95ER40896, a
Research Innovation Award and a Cottrell Scholar Award from Research
Corporation, and a Vilas Associate Award from the University of
Wisconsin.

\appendix
\section{Deformed Conifold Coordinates}\label{DefConcoord}
The embedding coordinates $\{z_i\}$ transform as ${\bf{4}}$ under the $SO(4)$ isometry group.
An alternative parametrization for the deformed conifold constraint (\ref{DefCon}) can also be given by a $2\times 2$ complex matrix:
\begin{equation}
W=\frac{1}{\sqrt{2}}\left(\begin{array}{cc}
z_3+iz_4&z_1-iz_2\\
z_1+iz_2&-z_3+iz_4
\end{array}\right)\equiv \left(\begin{array}{cc}
w_3&w_2\\
w_1&w_4
\end{array}\right)\label{defW}\,,
\end{equation}
defining equation for the conifold
can then be expressed in terms of $W$ as
\begin{equation}
\det W=w_3 w_4-w_1 w_2=-\frac{\epsilon^2}{2}\,.\label{DefCon2}
\end{equation}
The coordinates $\{w_i\}$ now transform as ${\bf(2,2)}$ under $SO(4)\sim SU(2)_{L}\times SU(2)_R$ isometry group.
The complex embedding coordinates of deformed conifold $\{z_1,z_2,z_3,z_4\}$ can be expressed in terms of the real coordinates $\{\rho\in{\mathbb{R}\,,~\psi\in[0,4\pi]\,,~\theta_{1,2}\in[0,\pi]\,,~\phi_{1,2}\in[0,2\pi]}\}$, $S=\rho+i\psi$ as:
\begin{eqnarray}\label{zdefconcoord}
z_1&=&\epsilon\left[\cosh\left(\frac{S}{2}\right)\cos\left(\frac{\theta_1+\theta_2}{2}\right)\cos\left(\frac{\phi_1+\phi_2}{2}\right)
+i\sinh\left(\frac{S}{2}\right)\cos\left(\frac{\theta_1-\theta_2}{2}\right)\sin\left(\frac{\phi_1+\phi_2}{2}\right)\right]\,,\nn\\
z_2&=&\epsilon\left[-\cosh\left(\frac{S}{2}\right)\cos\left(\frac{\theta_1+\theta_2}{2}\right)\sin\left(\frac{\phi_1+\phi_2}{2}\right)
+i\sinh\left(\frac{S}{2}\right)\cos\left(\frac{\theta_1-\theta_2}{2}\right)\cos\left(\frac{\phi_1+\phi_2}{2}\right)\right]\,,\nn\\
z_3&=&\epsilon\left[-\cosh\left(\frac{S}{2}\right)\sin\left(\frac{\theta_1+\theta_2}{2}\right)\cos\left(\frac{\phi_1-\phi_2}{2}\right)
+i\sinh\left(\frac{S}{2}\right)\sin\left(\frac{\theta_1-\theta_2}{2}\right)\sin\left(\frac{\phi_1-\phi_2}{2}\right)\right]\,,\nn\\
z_4&=&\epsilon\left[-\cosh\left(\frac{S}{2}\right)\sin\left(\frac{\theta_1+\theta_2}{2}\right)\sin\left(\frac{\phi_1-\phi_2}{2}\right)
-i\sinh\left(\frac{S}{2}\right)\sin\left(\frac{\theta_1-\theta_2}{2}\right)\cos\left(\frac{\phi_1-\phi_2}{2}\right)\right]\,.\nn\\
\end{eqnarray}
In the large radius limit $\rho\to \infty$, $\epsilon\cosh (\rho/2)\approx \epsilon\sinh (\rho/2)\approx \frac{r^{3/2}}{\sqrt{2}}$, the deformed conifold coordinates (\ref{zdefconcoord}) and (\ref{defW}) smoothly reduce to those of singular conifold.

\section{Generalized $z$-Embeddings}\label{SecGKuperembedding}
\paragraph{}
We next examine a generalization of the holomorphic embedding
(\ref{Kuperembedding}) proposed in \cite{Kuper}
\begin{equation}
z_4=f\left(z_1^2+z_2^2\right)\,,
\end{equation}
where $f(x)$ is an arbitrary function of $x=z_1^2+z_2^2=2(w_1w_2)$.
This class of embedding generally breaks the $SO(4)$ isometry group
down to the $U(1)$ which rotates $w_1$ and $w_2$ by opposite phases.
\paragraph{}
The general expression for the pull-back of NS-NS B-field in this
case is given by:
\begin{equation}
\heta_1=
\eta_4\vline_{z_4=f(x)}=B_{1\bar{1}}dz_1\wedge d\bz_1
+B_{1\bar{2}}dz_1\wedge d\bz_2+B_{2\bar{1}}dz_2\wedge d\bz_1+B_{2\bar{2}}dz_2\wedge d\bz_2\,,\label{DefBGkuper}
\end{equation}
with the various components given by the expression in
(\ref{KuperBfield}):
\begin{eqnarray}
B_{1\bar{1}}&=&2\left\{(z_3\bz_2-z_2\bz_3)(\bar{f}'\bz_1-f'z_1)+(f_2 z_1-\bar{f}_2\bz_1)(z_4\bz_2-z_2\bz_4)\right\}\nn\,,\\
B_{2\bar{2}}&=&-2\left\{(z_3\bz_1-z_1\bz_3)(\bar{f}'\bz_2-f'z_2)+(f_2 z_2-\bar{f}_2\bz_2)(z_4\bz_1-z_1\bz_4)\right\}\nn\,,\\
B_{1\bar{2}}&=&(z_4\bz_3-z_3\bz_4)+2\left\{(z_1(z_3\bz_1-z_1\bz_3)f'+\bz_2(z_3\bz_2-z_2\bz_3)\bar{f}')
- (z_3 \rightarrow z_4, f' \rightarrow f_2)
\right\}\,,\nn\\
B_{2\bar{1}}&=&(z_3\bz_4-z_4\bz_3)-2\left\{(\bz_1(z_3\bz_1-z_1\bz_3)\bar{f}'+z_2(z_3\bz_2-z_2\bz_3)f')
-(z_3 \rightarrow z_4, f' \rightarrow f_2)
\right\}\,.\nn
\\
\label{GKuperBfields}
\end{eqnarray}
Here
\begin{equation}
z_3=\sqrt{\epsilon^2-(x+f^2(x))}\,,~~~~f'(x)=\frac{d f(x)}{d x}~~~~f_2=-\frac{1+f'(x)}{2z_3}\,,
\end{equation}
the expression for the previous case (\ref{Kuperembedding}) is
recovered by setting $f'(x)=0$. The pull-back of the K\"ahler
two-form for the generalized Kuperstein embedding can be written as:
\begin{eqnarray}
\hat{\eta}_5=\eta_5\vline_{(z_4=f(x))}&=& J'_{1\bar{1}}\,dz_1\wedge d\bz_1+J'_{1\bar{2}}\,dz_1\wedge d\bz_2+J'_{2\bar{1}}\,dz_2\wedge d\bz_1+J'_{2\bar{2}}\,dz_2\wedge d\bz_2\,,\label{DefGKuperJp}\\
\hat{\eta}_4=\eta_4\vline_{(z_4=f(x))}&=& J''_{1\bar{1}}\,dz_1\wedge d\bz_1+J''_{1\bar{2}}\,dz_1\wedge d\bz_2+J''_{2\bar{1}}\,dz_2\wedge d\bz_1+J''_{2\bar{2}}\,dz_2\wedge d\bz_2\,.\label{DefGKuperJpp}
\end{eqnarray}
with the components
\begin{eqnarray}
J'_{1\bar{1}}&=&1+|z_1|^2 p(x)\,,~~J'_{2\bar{2}}=1+|z_2|^2 p(x)\,,~~J'_{1\bar{2}}=z_1\bz_2 p(x)\,,~~J'_{2\bar{1}}=z_2\bz_1 p(x)\,,\label{GKuperJps}\\
J''_{1\bar{1}}&=&S_1\bar{S}_1\,,~~J''_{2\bar{2}}=S_2\bar{S}_2\,,~~
J''_{1\bar{2}}=S_1\bar{S}_2\,,~~J''_{2\bar{1}}=S_2\bar{S}_1\,,\label{GKuperJpps}\\
p(x)&=&4|f'(x)|^2+\frac{|1+2f'(x)|^2}{|z_3|^2}\,,\label{Defkz}\\
S_1&=&\frac{1}{z_3}\left[(z_3\bz_1-z_1\bz_3)+2z_1(z_3\bz_4-\bz_3)f'(x)\right]  \\
S_2&=&\frac{1}{z_3}\left[(z_3\bz_2-z_2\bz_3)+2z_2(z_3\bz_4-\bz_3)f'(x)\right]
\end{eqnarray}
Notice here that these expressions reduces to the ones for simple
Kuperstein embedding (\ref{Kuperembedding}) when $f'(x)=0$. To check
the SUSY condition for the generalized embedding
(\ref{GKuperembedding}), we can first calculate
\begin{eqnarray}
\heta_5\wedge\heta_1&=&\frac{(z_1\bz_2-z_2\bz_1)}{|z_3|^2}
([\bz_3(1+z_3^2)f'(x)-z_3(1+\bz_3^2)\bar{f}'(x)]\nn\\
&+&(z_3\bz_4-z_4\bz_3)[2|f'(x)|^2(1+|z_3|^2)+f'(x)+\bar{f}'(x)])d\Omega\,,
\label{wedge51GKuper}
\end{eqnarray}
where $d\Omega=dz_1\wedge dz_2\wedge d\bz_1\wedge d\bz_2$. For an
arbitrary $f(x)$, the above expression does not vanish unless
$f'(x)=0$. One can also in principle calculate
$\heta_4\wedge\heta_1$, and demonstrate that it is non-vanishing for
arbitrary $f(x)$. In fact, one can consider a special case $f(x)=x$,
i.e. $z_4=z_1^2+z_2^2$; for each specific value of $z_4$, this is
identical to Karch-Katz embedding (\ref{KKembedding}), which we will
also show momentarily to be non-supersymmetric.
We thus conclude that the embedding (\ref{GKuperembedding}) cannot
be supersymmetric in the deformed conifold for arbitrary function
$f(x)$. However, with additional world volume flux, the
supersymmetric condition can still in principle be satisfied.

Let us remark also on a statement in \cite{Kuper} that the
$\kappa$-symmetry conditions must be satisfied because of a symmetry
argument.  The argument used the following logic.  Under the
interchange $z_1 \rightarrow z_2, z_2 \rightarrow z_1$, the Kahler
form $J$ and the embedding equation are invariant, while $B_2$ gets
a minus sign.  Thus $\hat{J} \wedge \hat{B}$ acquires a minus sign
under the interchange.  There is also a rotational symmetry (from
the $SO(3)$ preserved by the embedding equation) rotating $z_1$ and
$z_2$, under which $J$, $B$, and the embedding equation are all
invariant.  One can write $\hat{J} \wedge \hat{B} =
\phi(z_1,\bar{z_1},z_2,\bar{z_2}) dz_1\wedge dz_2\wedge d\bz_1\wedge
d\bz_2$.  Then $\phi$ is invariant under the rotation but acquires a
minus sign on the interchange, and $\cite{Kuper}$ claimed that the
only function which could satisfy both properties is $\phi=0$.  This
claim is not correct, as $z_1 \bz_2 - z_2 \bz_1$ is a
counterexample.

\section{Karch-Katz Embedding}\label{SecKKembedding}
\paragraph{}
We consider next the holomorphic embedding given in \cite{Karch}
\begin{equation}
w_1w_2=\frac{z_1^2+z_2^2}{2}=\mu^2\,,~~~~\mu\in{\mathbb{C}}\,,
\end{equation}
which explicitly breaks the $SO(4)$ isometry group into $SO(2)\times
SO(2)$, and these $SO(2)$ subgroups act by rotating the phases of
the ratios $w_1/w_2$ and $w_3/w_4$. In the asymptotic limit
$r^{3}\gg |\mu|^2$ limit, the embedding (\ref{KKembedding}) can in
fact be viewed as two copies of the D7 brane $w$-embedding we have
earlier (\ref{OYembedding}) $w_1=0$ and $w_2=0$, and they intersect
and fuse together at some finite radius by turning on D7-D7
interactions.
\paragraph{}
The pull-back of the NS-NS two form field in this case is given by
calculating:
\begin{equation}
\hat{\eta}_1=\eta_1\vline_{(w_1w_2=\mu^2)}=B_{1\bar{1}}\,dz_1\wedge d\bz_1+B_{1\bar{3}}\,dz_1\wedge d\bz_3+B_{3\bar{1}}\,dz_3\wedge d\bz_1+B_{3\bar{3}}\,dz_3\wedge d\bz_3
\end{equation}
where the various components are given by: {\small
\begin{eqnarray}
B_{1\bar{1}}&=&|z_2|^{-2}(z_1\bz_2-z_2\bz_1)(z_3\bz_4-z_3\bz_4)\,,\nn\\
B_{1\bar{3}}&=&-\left\{(z_2\bz_4-z_4\bz_2)+(z_1\bz_4-z_4\bz_1)\frac{z_1}{z_2}
+(z_2\bz_3-z_3\bz_2)\frac{\bz_3}{\bz_4}+(z_1\bz_3-z_3\bz_1)\frac{z_1\bz_3}{z_2\bz_4}\right\}\,,\\
B_{3\bar{1}}&=&\left\{(z_2\bz_4-z_4\bz_2)+(z_1\bz_4-z_4\bz_1)\frac{\bz_1}{\bz_2}
+(z_2\bz_3-z_3\bz_2)\frac{z_3}{z_4}+(z_1\bz_3-z_3\bz_1)
\frac{\bz_1 z_3}{\bz_2z_4}\right\}\,,\\
B_{3\bar{3}}&=&|z_4|^{-2}(z_1\bz_2-z_2\bz_1)(z_3\bz_4-z_3\bz_4)\,,\label{KKBfields}
\end{eqnarray}}
with
\begin{equation}
z_2^2=2\mu^2-z_1^2\,,~~~~z_4^2=(\epsilon^2-2\mu^2)-z_3^2\,.
\end{equation}
The components for the pull-back of the K\"ahler form $J$ can also
be calculated from:
\begin{eqnarray}
\hat{\eta}_5=\eta_5\vline_{(w_1w_2=\mu^2)}&=& J'_{1\bar{1}}\,dz_1\wedge d\bz_1+J'_{3\bar{3}}\,dz_3\wedge d\bz_3\,,\label{DefKKJp}\\
\hat{\eta}_4=\eta_4\vline_{(w_1w_2=\mu^2)}&=& J''_{1\bar{1}}\,dz_1\wedge d\bz_1+J''_{1\bar{3}}\,dz_1\wedge d\bz_3+J''_{3\bar{1}}\,dz_3\wedge d\bz_1+J''_{3\bar{3}}\,dz_3\wedge d\bz_3\,.\label{DefKKJpp}
\end{eqnarray}
The various functions here are:
\begin{equation}
J'_{1\bar{1}}=1+\frac{|z_1|^2}{|z_2|^2}\,,~~~~J'_{3\bar{3}}=1+\frac{|z_3|^2}{|z_4|^2}\,,~~~~J'_{1\bar{3}}=J'_{\bar{3}1}=0\,.\label{KKJp}\\
\end{equation}
\begin{eqnarray}
J''_{1\bar{1}}&=&-\frac{(z_1\bz_2-\bz_1 z_2)^2}{|z_2|^2}\,,
~~~~J''_{3\bar{3}}=-\frac{(z_3\bz_4-\bz_3 z_4)^2}{|z_4|^2}\,,\nn\\
J''_{1\bar{3}}&=&-\frac{(z_1\bz_2-\bz_1 z_2)(z_3\bz_4-\bz_3 z_4)}{z_2\bz_4}\,,
~~~~J''_{3\bar{1}}=-\frac{(z_1\bz_2-\bz_1 z_2)(z_3\bz_4-\bz_3 z_4)}{z_4\bz_2}\,.\label{KKJpp}
\end{eqnarray}
The SUSY condition (\ref{cond3s}) can be checked by calculating the
wedge products in turns:
\begin{eqnarray}
\hat{\eta}_5\wedge\hat{\eta}_1&=&-\frac{(z_1\bz_2-z_2\bz_1)(z_3\bz_4-z_4\bz_3)}{|z_2|^2|z_4|^2}
\left(|z_1|^2+|z_2|^2+|z_3|^2+|z_4|^2\right)d\Omega\nn\,,
\label{KKpcond}
\end{eqnarray}
where $d\Omega=dz_1\wedge dz_3\wedge d\bz_1\wedge d\bz_3$. In
(\ref{KKpcond}) we have a product of positive definite sum with a
generically non-vanishing number
$-4{\rm{Im}}(z_1\bz_2){\rm{Im}}(z_3\bz_4)$. Similarly we calculate
$\hat{\eta}_4\wedge\hat{\eta}_1$ and obtain:
\begin{eqnarray}
\hat{\eta}_4\wedge\hat{\eta}_1&=&\frac{(z_1\bz_2-z_2\bz_1)(z_3\bz_4-z_4\bz_3)\sum_{i\neq j=1}^4(z_i\bz_j-z_j\bz_i)^2}{|z_2|^2|z_4|^2}d\Omega\nn\,.\label{KKppcond}
\end{eqnarray}
Once again we have a product of a positive sum and a non-vanishing
number. We can again show that the SUSY condition (\ref{cond3}) is
not satisfied for Karch-Katz embedding (\ref{KKembedding}) without
additional magnetic field $F_2$. This result is perhaps not so
surprising given the fact that Karch-Katz embedding can be regarded
as two copies of $w$-embeddings asymptotically, and we have shown
that $w$-embedding cannot be supersymmetric unless additional world
volume flux is turned on.

\section{Comparison with Benini's Proposal}
\label{appBenini}

Benini \cite{Benini} has proposed a form for the asymptotic flux in
the limit of large radius.  The purpose of this Appendix is to
compare our notation with his so that it is clear how our proposal
compares with his in the relevant limit.

In the large radius limit, we may take one of the angles $\theta_1$
or $\theta_2$ to vanish.  A helpful piece of intuition is that the
worldvolume of the D7-brane in this limit ``splits" into two
branches, corresponding to one or the other $\theta_i$ vanishing.

Let us take the limit where one of the angles, say $\theta_2$, is
small, and then work order-by-order in $\theta_2$. The D7-brane
worldvolume in the compact directions can be described by the
coordinates $\theta_1,\theta_2,\phi_1,\phi_2$. However, to compare
with the large radius limit we will trade $\theta_2$ and $\phi_2$
for $r$ and $\psi$.

In this limit, the (1,1) basis forms $P$ and $Q$ take the form
\beq
P&\sim& \left[\frac32 \Omega_{11} + 3\frac{dr}{r}\wedge g^5 \right]+
\cot\frac{\theta_1}{2}(d\theta_1\wedge g^5 -3 \frac{dr}{r}
\wedge \sin\theta_1 d\phi_1) \nonumber \\
 Q
&\sim& -\frac12 \tan\frac{\theta_1}{2}(d\theta_1\wedge g^5 -3
\frac{dr}{r}
\wedge \sin\theta_1 d\phi_1)
\label{PQBeninilim}
\eeq
where $g^5 = d\psi + \cos\theta_1 d\phi_1 + \cos \theta_2 d\phi_2$.
The claim in \cite{Benini} was that the term in square brackets in
$P$ was a solution to the inhomogeneous Bianchi identity in the
large radius limit. This corresponds to a combination of $P$ and $Q$
with $\alpha =2k$ and $\beta =2k\cot^2\frac{\theta_1}{2}$, or in
other words
\beq
\hcF= 2k P +\left( \frac{2k}{\sin^2\frac{\theta_1}{2}}-2k \right) Q
\label{Beninilim}
\eeq
which we have written in a form to make clear the comparison with
formulas such as (\ref{b1}).

If one took the large radius limit on the other branch, with
$\theta_1$ small, the form of $\hcF$ is almost the same, with
$\theta_1$ and $\phi_1$ replaced by $\theta_2$ and $\phi_2$
everywhere, and multiplication by an overall factor of $-1$ (due to
the antisymmetry of the basis two-form $\omega_2$ which appears in
the background 2-form potential $B_2$.)  In our formalism, this sign
flip is built into the basis forms $P$ and $Q$ which are
antisymmetric under the interchange of indices $1 \leftrightarrow
2$, while the functions $\alpha$ and $\beta$ are symmetric.

Finally, note that our asymptotic form for $\alpha$ in the large
radius limit (\ref{alphalimsmallt2}) in the radial coordinates is
\beq
\alpha = -2k \log\log r + {\rm const} + O(1/\log r)
\eeq
with a leading $\log \log$ term; this term is effectively constant
in the leading $1/\log r$ approximation, but it does affect the
field equations at subleading order (and therefore has to be
included), as we argued in Section \ref{subleading}.

\end{document}